\documentclass[sigconf]{acmart}

\setcopyright{none} % No copyright notice required for submissions
\acmConference[Anonymous Submission to ACM CCS 2021]{ACM Conference on Computer and Communications Security}{Due 15 May 2019}{London, TBD}
\acmYear{2021}

\author{Junpeng Wan, Yanxiang Bi, Zhe Zhou}\affiliation{Fudan University}\email{zhouzhe@fudan.edu.cn}

\author{Zhou Li}\affiliation{University of California, Irvine}\email{zhou.li@uci.edu}

\usepackage{graphicx}
\usepackage{xcolor}
\usepackage{listings}

\usepackage{subcaption}
\usepackage[ruled,vlined]{algorithm2e}
\usepackage{multirow}
\usepackage{makecell}
\usepackage{colortbl}
\usepackage{url}

\usepackage{booktabs}

\newcommand{\etal}{{\textit{et al.}}}
\newcommand{\ie}{{\textit{i}.\textit{e}.}}
\newcommand{\eg}{{\textit{e}.\textit{g}.}}
\newcommand{\etc}{\textit{etc}.}
\newcommand{\ignore}[1]{}
\newcommand{\zztitle}[1]{\vspace{2pt}\noindent\textbf{#1.}}
\newcommand\zl[1]{{\color{magenta}{\textbf{\{zl: {\em#1}\}}}}}
\newcommand{\attack}{\textsc{Volcano}}
\usepackage{url}

\begin{document}
\title{Volcano: Stateless Cache Side-channel Attack by Exploiting Mesh Interconnect}

\begin{abstract}
Cache side-channel attacks lead to severe security threats to the settings that a CPU is shared across users, e.g., in the cloud.
The existing attacks rely on sensing the micro-architectural state changes made by victims, and this assumption can be invalidated by combining spatial (\eg, Intel CAT) and temporal isolation (\eg, time protection).
In this work, we advance the state of cache side-channel attacks by showing stateless cache side-channel attacks that cannot be defeated by both spatial and temporal isolation.

This side-channel exploits the timing difference resulted from interconnect congestion. Specifically, to complete cache transactions, for Intel CPUs, cache lines would travel across cores via the CPU mesh interconnect. Nonetheless, the mesh links are shared by all cores, and cache isolation does not segregate the traffic. An attacker can generate interconnect traffic to contend with the victim's on a mesh link, hoping that extra delay will be measured.
With the variant delays, the attacker can deduce the memory access pattern of a victim program, and infer its sensitive data.
Based on this idea, we implement \attack{} and test it against the existing RSA implementations of JDK. We found the RSA private key used by a victim process can be partially recovered. In the end, we propose a few directions for defense and call for the attention of the security community.

%According to the side channel, we design \attack{}, where the attacker produces huge mesh traffic and measures the delay owing to victim cache activities. With the delay sequence, the attacker is able to learn very sensitive information about the victim.

%When cache is isolated between OS's or VMM's boundary, attacker's cache lines would no longer be interfered by victim's cache activities, preventing potential cache attacks.
%The attack does not recognize newly inserted or evicted cache lines by measuring the timing of memory access. Instead,

\end{abstract}

\maketitle

\section{Introduction}

%why cache side-channel matters
Isolation is one of the fundamental security principles to protect sensitive information. As a concrete implementation, virtual memory isolation, which separates the memory space between processes managed by an operating system (OS), and virtual machines (VMs) managed by a hypervisor, is universally deployed.
Yet, isolation at the memory level is not sufficient, as there are many other shared hardware resources that can be exploited. One prominent example is CPU cache. Since the memory access also leads to updates on cache but memory isolation is not directly mapped to the cache, a plethora of previous works exploit the shared cache as a side-channel to carry out timing attacks, and various sensitive information has been found unprotected from the lens of cache, resulting in the leakage of user input~\cite{wang2019unveiling}, cryptographic key~\cite{tromer2010efficient}, \etc

%Prime+Probe, all shared cache, cache isolation
\zztitle{Existing Cache Side-channel Attacks and Defenses} 
%It allows attackers to learn the memory access pattern of a co-located victim core and then infer sensitive information about the victim. 
By timing the interval of accessing a cache line, existing attacks learn whether the associated memory addresses have been loaded by a victim program, and further deduce the sensitive information. Many techniques have been proposed, like Flush+Reload~\cite{yarom2014flush+}, Prime+Probe~\cite{disselkoen2017prime+}, and the recent Xlate~\cite{van2018malicious}, based on different assumptions, \eg, using shared libraries~\cite{yarom2014flush+},  sharing LLC (Last-level Cache) across cores~\cite{liu2015last}, and sharing MMU (Memory Management Units), \etc\ 

%effects made by victim to attacker's cache remain, unless destroyed by another cache activity.

%is present in the shared cache. By measuring the time spent on accessing a piece of memory, the attacker knows if the cache line is missed or hit. 
%Thereby, an attacker can reason sensitive information according to the knowledge that if the cache line presents.

These attacks can be categorized to \textit{stateful cache attacks}, as the victim program introduces micro-architectural state changes that can be sensed by an attacker~\cite{ge2019time}, \eg, through creating eviction sets.
%Specifically, no matter what assumption is made by those attacks, they rely on a spot that attackers can evict victim's cache lines, or vice verse, which can be called eviction set conflict.
However, this attack condition may not be fulfilled nowadays, due to rise of spatial and temporal isolation. For example, Intel Xeon CPUs introduces \textit{Cache  Allocation  Technique (CAT)}~\cite{cat}, which was designed to maintain QoS of cache usage, but later found to be a panacea for cache attacks~\cite{liu2016catalyst}. CAT assigns LLC cache ways to cores exclusively, which spatially isolate LLC and break all attacks based on eviction set conflict on LLC (\eg, ~\cite{liu2015last}).  Academic proposals extend the protection realm to other cache units, like directories~\cite{yan2019secdir}. Besides spatial protections, \cite{ge2019time} proposed an OS abstraction for temporal isolation, which is effective against nearly every cache attack with the existing hardware support.

%Thereby, the research question we want to find an answer is: \textit{is cache isolation the silver bullet against any cache side-channel attack?}

%\zl{~\cite{liu2016catalyst} shows CAT is not for isolation, we need to fix the claim. Intel claim CAT as a kind of isolation in the webpage of CAT;Done}
%Cache isolation can well defeat non-volatile cache attacks. Non-volatile cache attacks rely on the cache co-location between the victim and the attacker. The attacker's cache lines can be interfered (\ie either evicted or pulled up) by the victim only when they co-locate in the same cache set. Cache isolation techniques, , with which there is no longer cache co-location between the attacker and the victim. Thereby, non-volatile cache attacks do not work under the protection of cache isolation. 
%Recent cache attacks can even work on non-inclusive cache but can still be defeated by cache isolation~\cite{yan2019attack}.

%New opportunities brought by mesh 
\zztitle{Stateless Cache Side-channel Attack} 
Given that spatial and temporal partition can eliminate the root cause of cache side-channel attacks, or micro-architectural state change, \textbf{is it possible to evolve the attack to make it \textit{stateless}?} It sounds counter-intuitive but there is hope. According to Ge \etal~\cite{ge2019time}, there exist a few stateless channels, \eg, on memory bus contention~\cite{wu2014whispers}, that do not incur micro-architectural state change. These stateless channels were uncovered a long time ago~\cite{wu2014whispers}, but they are all limited to construct covert channels~\cite{ge2019time}. If one can use the stateless channel as the medium to attack the cache, the goal might be achievable. A more important feature is that even the highly secured scheme like temporal isolation is admittedly ineffective against stateless channels, due to the lack of hardware support on bandwidth partition~\cite{ge2019time}, indicating that stateless attacks could hardly be defeated.

%Previous work~\cite{ge2019time} believes that stateless channels can hardly leak sensitive information, because literature either only applies to covert channel~\cite{wu2014whispers} or only simulated~\cite{wang2012efficient}. At the same time, ~\cite{ge2019time} admits that stateless channels can hardly be mitigated, due to lack of isolation. 
%``as long as the interconnect does not leak data or address information, they are probably infeasible.''~\cite{ge2019time}

%\texttt{Are stateless channels harmless?} Our answer is definitely NO! 
In this work, we revisit the direction of stateless channels and explore the possibilities of combining them with cache attacks. By investigating the latest architecture of Intel CPU, \eg, Xeon Scalable Processor~\cite{xeon}, we identified a new stateless channel, and found it can be exploited for our goal.
%attacks can indeed make devastating consequences, like RSA key leakage.
In these CPUs, the interconnect between cores and other units (\eg, I/O units) is turned into a \textit{mesh} network, consisting of a 2D matrix of \textit{tiles}. 
%Besides, LLC is partitioned into slices shared across cores. Each core, LLC, or other unit is enclosed in a tile and two adjacent tiles share a mesh link. 
Figure~\ref{fig:mesh_ring} right shows the mesh network. 
Although mesh interconnect shows great advantage in latency and bandwidth on multi-core CPUs~\cite{bell2008tile64,vangal200780}, it could leak information about the cache transactions of a program, because different cache transactions share the mesh links. Hence, there exists bandwidth contention when cache lines run across mesh, and the attacker might use that to infer the victim's secret.
%intentionally cause mesh congestion, and time the cache access to infer the victim's secret.

%Though cache isolation protects cache lines at still, cache lines \textit{on the move} fall out of its protection scope.

%The mesh interconnect is responsible for all the message exchanges between cores and LLC slices. When a core running victim thread issues a cache transaction to LLC, the core and the responsible LLC slice exchange messages and cache lines via the mesh network, causing traffics on a mesh path from the requesting core to the LLC slice. 

%The side channel stems from the congestion happening on the interconnect between geographically disaggregated last level cache (LLC) slices.
%Different from L1 and L2 caches, LLC is shared across all cores in recent Intel CPUs. 
%To provide higher throughput, modern LLC is split into multiple slices that can concurrently serve all the cores
%The LLC slices and cores are interconnected with a 2D network named mesh, which shows 

%We firstly reveal a kind of cache side channel that does not require cache conflict between the victim and the attacker. 

Based on this insight, we propose stateless cache side-channel attacks, or \attack. Our key idea is to let a core occupied by an attacker program keeps probing the path that the cache transactions of a victim program pass by, and measure the interval. When the core occupied by the victim program accesses a remote LLC slice, the accumulated mesh traffic volume will rise, hence increasing the interval observed by the attacker, especially when the mesh link is congested. By probing the mesh link at high frequency, the attacker could reconstruct the execution flow of the victim program, and further deduce the secret.

\zztitle{Challenges} Although \attack{} is easy to understand, implementing an effective attack is challenging. Here we list the key challenges.

\begin{itemize}
    \item \textbf{Locating the traffics on the CPU layout}. Though the attacker and the victim threads run within the same CPU, the attacker does not know where the victim traffic originates from and to.
    
    \item \textbf{Synthesizing targeted probe traffic.} The mesh interconnect is invisible to the upper-layer application, and there is no API to let an attacker direct mesh traffic to a given destination.
    
    \item \textbf{Sampling a mesh link.} Though Intel provides tools like PMON~\cite{pmu_manual} to measure mesh traffic for each mesh link, we found they require root privilege and are unable to sample a mesh link at a high rate. 

    \item \textbf{The coarse-grained probe.} \attack{} can only observe whether there are traffics from/to a core rather than the status of a specific cache line.
\end{itemize}

\zztitle{Attack Techniques}
To tackle the above challenges, we design a new method to reverse engineer the CPU layout and map the core/CHA (Cache Home Agent) IDs to all the tiles. Based on the layout, we carry out a theoretic analysis of the distribution of mesh traffic from/to a core, to guide the selection of mesh links to be contended with. To enable targeted and robust probes, we carefully construct an eviction set, which can be mapped to a specific LLC slice, and contained in a L2 set. Accessing the eviction set generates stable mesh traffics and the attacker can rely on the timing to measure the congestion on mesh. \textbf{To notice is that the eviction set does not conflict with the victim.} It is only used to trigger mesh traffics. Therefore, defenses trying to prevent adversarial cache eviction can be evaded under \attack{}.

%Then, the attacker can precisely know the location of victim core on mesh.
%We analyze various kinds of potential victim mesh traffic and its distribution over mesh tiles. Then the attacker is able to precisely co-locate with the victim traffic.

As a showcase for the attack effectiveness, we analyzed fast modular exponentiation algorithm and sliding windows algorithm of RSA respectively, by using \attack{} to recover the 2048-bit private RSA key. For the fast modular exponentiation algorithm, we found the attacker has \textbf{47\%} chances to recover \textbf{2040 bits correctly}. For the sliding windows algorithm, which was included in the crypto library of JDK, the attacker can recover over 30\% of the 2048 bits, with the help of a cryptographic method~\cite{bernstein2017sliding}. 

\zztitle{Contributions} We summarize the main contribution of our work as follows.
\begin{itemize}
  \item We identify a new security implication of CPU mesh interconnect, and show it can be exploited to construct a powerful stateless side-channel.
  %firstly identify that the timing difference of traffic over interconnect reveal cache activities on the co-located mesh path, which constitutes a timing side channel. 
  \item We propose a new method to reverse-engineer the CPU layout, which allows an attacker to precisely locate the victim core on the mesh structure and analyze the traffic on mesh. 
  %Besides, we analyze various mesh traffics and its path, so the attacker can even co-locate with the victim traffic.
  \item We develop \attack{}, using the stateless channel as a probe to conduct cache side-channel attack. 
  
  %with the volatile cache side channel. To show the practicality of volatile cache side channel, we design schemes to produce mesh traffic that co-locates with the victim traffic. By measuring the timing difference, \attack{} allows the attacker to learn senstive information about the victim.
  \item To show the consequences of \attack, we evaluate our attack on the off-the-shelf RSA implementations.
  %\item We propose a few directions for defense. 
\end{itemize}

\section{Background and Related Works}
\label{sec:background}

In this work, we investigate the security of the cache architecture of Intel Xeon Scalable Processors~\cite{xeon}, which have gained a prominent market share in cloud computing~\cite{intel_share}. We first overview their cache design. Then, we briefly introduce the prior cache side-channel attacks and the corresponding defenses. Finally, we overview the research of stateless channels that serve as our attack primitives.

\subsection{Cache Architecture of Intel Xeon Scalable Processors}
\label{subsec:cache}

%We introduce the technique update related with our work in recent Intel Xeon scalable CPU family CPUs.

%talk about interconnect, and the 

Though at the high level, the Intel Xeon Scalable Processors inherit the L1/L2/L3 (LLC) cache hierarchy from their predecessors, cache management is upgraded under mesh interconnect. We introduce this feature first. Then we describe the general cache hierarchy and how it is evolved.

\begin{figure}[ht]
    \includegraphics[width=0.45\textwidth]{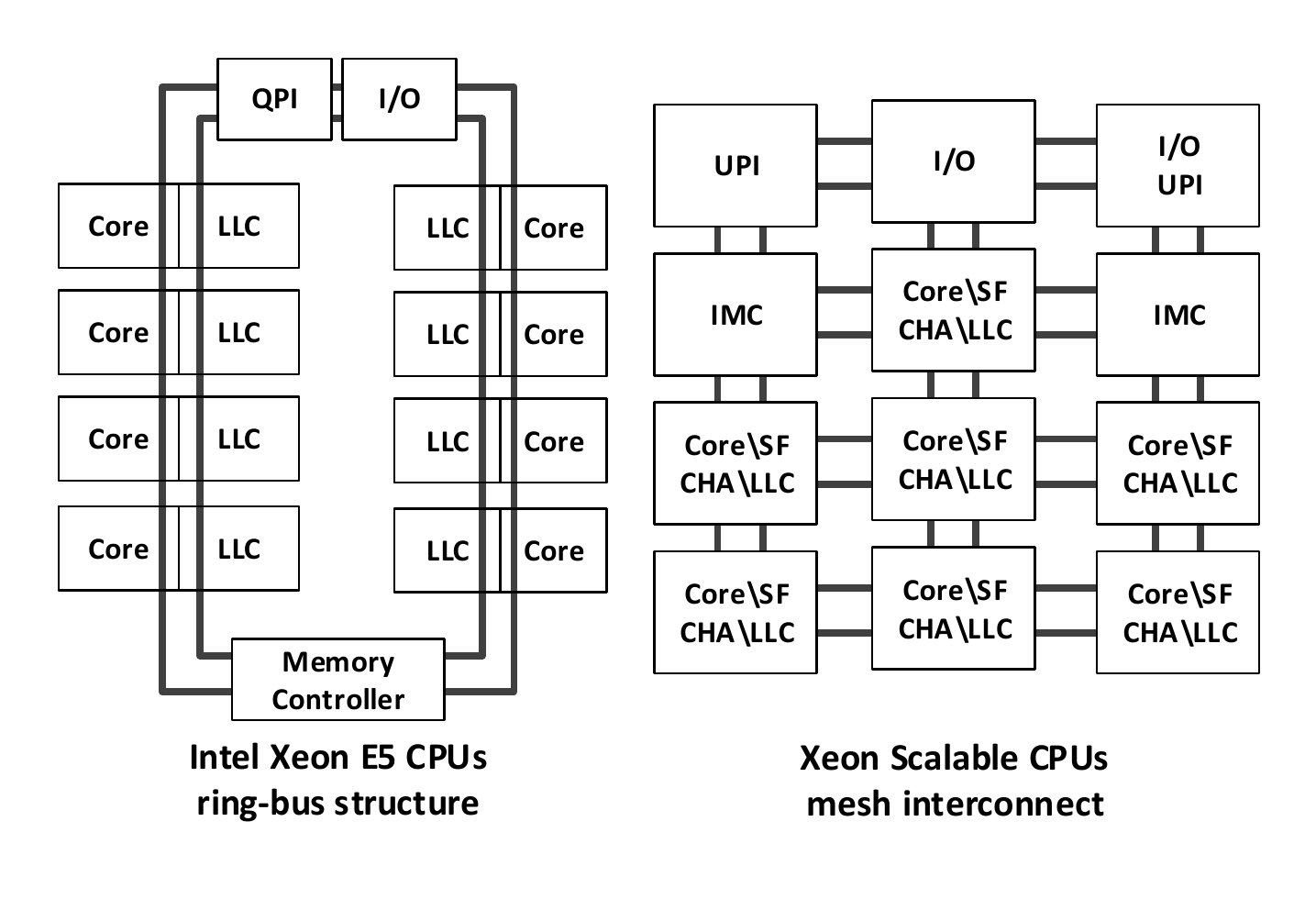}
    \caption{Comparison between ring-bus and mesh structures.     Not all cores/components are drawn to save space.
    %\zl{can you add an example about how data are routed from one core to another to the figures?}
    }
    \label{fig:mesh_ring}
\end{figure}

\zztitle{Mesh Interconnect}
When a CPU chip contains multiple cores, the connection topology among them, or \textit{on-chip interconnect} is a key factor determining the CPU performance. The old design of interconnect (\eg, in Intel Xeon E5) mimics the multiprocessor architecture, in that a shared \textit{ring-bus} connects all cores together ~\cite{gilbert2012intel}, as illustrated in Figure~\ref{fig:mesh_ring} left. 
However, the core-to-core latency could increase linearly along with the growth of cores within one CPU die, because the communication between two cores could be routed through all other cores. 

Since Xeon Skylake-SP server CPU family~\cite{mesh_wiki} (released in 2017), Intel revamped the interconnect design with \textit{mesh}, which is also adopted by the latest generation of Intel server CPUs, \eg, Xeon Cooper Lake-SP, and expected to be the default design in the near future~\cite{icelake-sp}.
Besides Intel CPUs, mesh interconnect has also been adopted by other processors, like Tile Processors~\cite{bell2008tile64,taylor200316,vangal200780}, and ARM server CPUs~\cite{arm_mesh}.
%\zl{can you find which AMD supports mesh? I find an ARM CPU with mesh. Currently AMD uses another structure.}. 
In essence, the chip is structured as a 2D matrix of identical \textit{tiles}~\cite{wentzlaff2007chip} and each tile either consists of a core (together with cache), or a non-core component like \textit{IMC (Integrated Memory Controller)}, \textit{UPI (Ultra Path Interconnect)}, and I/O unit. Each tile is connected to its vertical and horizontal neighbors, and the traffic of each direction (in total 4) is managed through a \textit{mesh stop} inside the tile. Figure~\ref{fig:mesh_ring} right illustrates the mesh structure.
Mesh interconnect caps the core-to-core latency at a much lower rate, because the number of hops between any pair of tiles is only proportional to the square root of the number of tiles. 
%Mesh interconnect allows more cores in a single CPU. 
%To provide rapid response in terms of inter-core connection, 
%Intel introduced the 2D mesh architecture. 
%As for the reason that mesh outperforms ring-bus in terms of inter-core connection latency and bandwidth, the number of hops between the farthest core pair increases linearly against the increasing number of cores (see Figure~\ref{fig:mesh_ring}). 
%Therefore, the latency between cores decreases. 
In addition to reducing latency, mesh interconnect also enlarges the bandwidth available to each core, because the communication traffic is distributed across more tiles and less likely to congest a route.
%A core in ring-bus only has two outgoing links, but a core in mesh has four, which balances the interconnect traffic among cores. 
%\zl{change switch to mesh stop, and highlight it throughout paper}
%Cores, as well as non-core components including I/O ports, memory controllers and inter-socket links, are all connected into mesh. 

%Mesh acts as a link-layer network switch that carries different kinds of traffics originated from those different tiles. 

It is worth mentioning that the mesh of Intel chips implements a simple YX routing protocol,  as such the route is \textit{always the same} given a pair of source and a destination tiles~\cite{mesh_wiki}. 
Specifically, a packet always flows to the row of the destination tile \textit{vertically}, and then moves to the destination \textit{horizontally}. 
This static routing mechanism simplifies the implementation of mesh, but on the other hand opens up new cache side-channels, when the adversary carefully crafts cache requests, and we elaborate the attack method in Section~\ref{sec:probe}.

%does not require each tile to care about the traffic dynamics on mesh, and also . 

\zztitle{Cache Hierarchy}
%general cache hierachy, use mengjia's text
Modern processors all feature a hierarchy of cache to localize the frequently accessed data and code, in order to reduce access latency. Each core has its own private cache, \eg, L1 and L2 cache. And there is also L3 cache, or \textit{last-level cache (LLC)}. From L1 to LLC, the access speed is reduced but the size allocated per core is increased. A cache is organized in $S$ sets, and each set has $W$ ways. Each read/write access touches a cache line, which occupies one way of cache. The cache is usually \textit{set-associative}, meaning that a memory block could be placed into any of the $W$ ways within a set.

%inclusive, non-inclusive, 
As mesh interconnect leads to higher bandwidth and lower latency for core-to-core communications, Xeon Scalable Processors breaks LLC into slices and evenly distribute them among cores. 
%The LLC of skylake-SP is designed in distributed fashion. 
As shown in Figure~\ref{fig:mesh_ring} right, each tile consists of a core (with L1 and L2 embedded), a LLC slice, a \textit{Snoop Filter (SF)} and a \textit{Cache Home Agent (CHA)}. 
%Each core has its private cache (\ie L1 and L2) that is exclusively used by the core. The LLC slices, however, are shared by all the cores in the die. 
Table~\ref{tab:cache_config} lists the cache configurations of the Xeon Scalable CPU families.

%\zl{add citations, RA; inclusive seem not to be related to our work.} 

How a memory address maps to a cache line depends on the set index within the address bits and the mapping algorithm. We illustrate the structure under Xeon Scalable Processors in Figure~\ref{fig:address} in Appendix, summarized from~\cite{farshin2019make, yan2019attack}.

\ignore{
\zztitle{UPI}

CHA also resolve cache coherence between CPUs, so that peer CPUs in
}

\subsection{Cache Side-channel Attack}
\label{subsec:cacheattack}

%\zl{to be fixed with ~\cite{ge2019time} and ~\cite{van2018malicious}}
Cache side-channel attack bypasses memory isolation and 
%, and poses severe threats to the confidentiality of critical data.
it is particularly concerning in the cloud setting, where multiple users share the same physical machines 
%and even the strong isolation through virtual machines does not prevent such attacks
~\cite{liu2015last, zhang2014cross, irazoqui2015s}.
Below we overview the existing attack methods, and classifies them by whether they assume memory sharing between victim and attacker. The overview is not meant to be exhaustive and the attacks can be characterized in other ways, e.g., whether a core is shared, and we refer the interested readers to surveys like~\cite{ge2018survey}.

%which can be launched in many CPU sharing scenarios and poses great threat to users, like RSA decryption leakage. 

%At the high level, this attack assumes the attacker is able to share the same cache (\eg, LLC) with the victim, so when a victim accesses a piece of code and data valuable to the attacker, causing changes to the cache lines, it can be inferred from the timing analysis. 
%Below we describe two primary attack strategies. Though our attack uses some primitives (\eg, eviction set) of the prior attacks, it relies on very different assumptions.

%We summarize some related non-volatile cache attacks as follow.

\zztitle{Sharing Memory}
Running processes often share identical memory pages, e.g., through the shared libraries, to reduce memory overhead. Shared memory leads to shared cache, and \textsc{Flush+Reload} exploits such condition for cache side-channel attack.
%Flush+Reload attack assumes the attacker and victim share memory. \eg, through shared libraries, with the victim. 
It takes three steps. First, the attacker sets all the cache lines mapped to the shared memory as invalid, by using cache clearance instruction \texttt{clflush}. Then, the attacker waits a period of time for the victim to access the addresses of the shared memory. Finally, the attacker accesses the shared addresses and counts the cycles (\eg, through \texttt{rdtsc}) to measure the latency, and infer the code/data access pattern of the victim.
%According to the measurement, she can infer which lines are hit (faster) and which lines are missed (slower), so the access pattern of the victim is derived.

%Further she learns which lines have been accessed by the victim, as the hit lines are in cache because the victim has just accessed the line.

\textsc{Flush+Reload} has been demonstrated effective on LLC~\cite{yarom2014flush+} of a PC and cloud instances~\cite{zhang2014cross}, resulting in leakage of encryption keys~\cite{irazoqui2015s} and keystroke events~\cite{wang2019unveiling}.
It has been evolved to variations~\cite{ge2018survey} like \textsc{Flush+Flush} ~\cite{gruss2016flush+}, which is stealthier by avoiding the extra memory access. On the other hand, this attack can be mitigated when \texttt{clflush} is banned~\cite{ban_clflush}. 
To address this limitation,  \textsc{Evict+Reload}~\cite{gruss2015cache} was proposed, which uses cache conflicts as a replacement for \texttt{clflush}.

%\zztitle{Evict-Reload} \zl{write sth?}

%\textbf{Prime+Probe}
\zztitle{Not Sharing Memory}
When memory is not shared, an attacker can still cause cache contention, due to memory addresses of different programs can share a cache set.
%Different from Flush+Reload, Prime+Probe does not rely on the shared memory between the attacker and victim. 
\textsc{Prime+Probe} exploits such feature, and it also takes three steps. First, the attacker collects a set of cache lines that can fill a cache set and access the related memory addresses.
%evicts victim's cache lines by creating conflicts on the cache set where the victim's addresses are mapped to. 
%When the attacker and the victim are in different cores, the private cache of the victim has to be evicted as well. 
Next, the attacker waits for the victim to evict the cache lines. Finally, the attacker measures the access latency.
%(\eg, through \texttt{rdtsc}), so whether the cache is evicted by the victim is learnt.

Though \textsc{Prime+Probe} initially targets L1 cache~\cite{osvik2006cache}, LLC that is inclusive has also been attacked~\cite{irazoqui2015s, liu2015last}. 
A number of variations have been developed~\cite{ge2018survey}. For instance,  \textsc{Prime+Abort}~\cite{disselkoen2017prime+} measures the Intel TSX (Transactional Synchronization Extensions) abort rather than access latency. Instead of letting the victim evict the cache lines, \textsc{Evict+Time} let the attacker evict a cache set, and then invokes the victim operation~\cite{lawson2009side, gras2017aslr}. 

%which has done the attack over the last level cache by exploiting the inclusiveness. Another representative prime probe could be~\cite{irazoqui2015s}, which break the VM sandboxing via prime-probe cache attacks.

%\zl{need to have more citations.}
%\zztitle{Transient Execution Vulnerabilities}

\zztitle{Indirect Attacks}
Recently, researchers started to investigate the interplay between other CPU units and cache, to make the attack more evasive. For instance, \textsc{Xlate}~\cite{van2018malicious} and 
\textsc{TLBleed}~\cite{gras2018translation} exploited MMU (Memory Management Units) and TLB (Translation Lookaside Buffers) to leak victim's cache activity. The recent Intel Xeon Scalable Processors started to use non-inclusive LLC, which raised the bar for LLC cache attacks.
%Most of the existing LLC cache attacks assume the LLC is inclusive, so the attacker can evict cache lines in the victim's private cache. However, when LLC is non-inclusive, the attacker lacks visibility to the mapping between LLC to the private cache, making the LLC cache attacks to be difficult to execute. 
Yet, Yan \etal~\cite{yan2019attack} showed that by targeting cache \textit{directories} (or Snoop Filter), the units tracking which core contains a copy of a cache line, attacking non-inclusive LLC is feasible.
%The fundamental issue underlying this type of attack is that the directories are not isolated.

One major assumption of the prior attacks is that the attacker's code is on the same machine as the victim's.
%\zztitle{Remote cache side-channel attacks}
Recently, attacks over network connections were studied. By exploiting RDMA (Remote Direct Memory Access) and DDIO (Data Direct I/O), a remote attacker can access LLC~\cite{tsai2019pythia} of CPU and cache inside NIC~\cite{kurth2020netcat}, launching side-channel attacks. 
%\attack{} is local as it does not use RDMA, but it might be possible to be combined with RDMA to make the attack remote.

On an orthogonal direction, transient execution attacks~\cite{canella2019systematic} like \textsc{Spectre}~\cite{kocher2019spectre}, \textsc{Meltdown}~\cite{lipp2018meltdown} and \textsc{Foreshadow}~\cite{van2018foreshadow} modulate the state of the cache to construct \textit{covert channels}, and exfiltrate information from speculatively executed instructions. \attack{} focuses on side channels and we will investigate whether \attack{} can be leveraged by transient execution attacks in the future.

%Recent cache attacks like Spectre~\cite{kocher2019spectre} and Meltdown~\cite{lipp2018meltdown} exploited the transient execution feature of CPUs, like Out-of-Order Execution, to construct side-channels. Following up Spectra and Meltdown are many new variants, like RIDL~\cite{van2019ridl}, which targets the data in-flight buffer. In essence, transient attacks construct transient instructions to bypass the data/code boundary of a program, \eg, through branch mispredictions, and then leverage covert-channels like the shared cache to exfiltrate the sensitive the information. 
%Canella \etal~\cite{canella2019systematic} systematically evaluated the attack variations and defenses under this attack type. \attack{} differs from the transient attacks in that no transient instruction is leveraged.

\begin{table*}[ht]
    \centering
    \begin{tabular}{c|cccccl}
         & \makecell{Co-location\\Assumption} & \makecell{Attack\\Channel} & \makecell{Page \\ Coloring} & \makecell{Hardware \\ Isolation} & \makecell{Temporal\\Isolation} & Key Feature\\ \hline
         Flush+Reload~\cite{yarom2014flush+} & Memory & LLC & \checkmark & CAT & \checkmark & Exploit shared memory \\ \hline
         Prime+Probe~\cite{irazoqui2015s, liu2015last} & Cache & LLC & \checkmark & CAT & \checkmark & No need to share memory \\ \hline
        TLBleed\cite{gras2018translation} & Core & TLB & - & Disable HT  & \checkmark & Attack TLB not cache \\ \hline
        \makecell{Attack Directories,\\ not caches \cite{yan2019attack}} & Cache & Directory & \checkmark & SecDir~\cite{yan2019secdir} & \checkmark & Work on non-inclusive cache \\ \hline
        Prime+Abort\cite{disselkoen2017prime+} & Cache & TSX Status & \checkmark & CAT & \checkmark & Does not rely on timing instruction \\ \hline
        Xlate\cite{van2018malicious} & Cache & MMU & - & - & \checkmark & Lure MMU to access cache \\ \hline
        \attack{} & Mesh &Mesh &-&-&-& No need to co-locate in cache
    \end{tabular}
    \caption{Comparison of cache side-channel attacks. ``\checkmark'' means the attack can be defended. ``-'' means the defense is ineffective. }
    \label{tab:model_compare}
\end{table*}
%\footnotesize{ $^1$\cite{ge2019time} assumes all critical shared resources should be partitioned. However, directory was not known as vulnerable until~\cite{yan2019attack}. Therefore, current \cite{ge2019time} could not defeat \cite{yan2019attack}, but they could if taken into consideration~\cite{yan2019secdir}. }
\subsection{Defenses against Cache Side-channels}
\label{subsec:isolation}

%\zl{to be edited}
%\zl{downlplay breaking cache isolation}
%\zl{use section 4 of ~\cite{van2018malicious} for defense, there are a lot more we haven't surveyed.}
%\zl{software, hardware, and ~\cite{ge2019time}}

\zztitle{Software Defenses} Since OS controls the allocation of memory to programs, the access to the physically-indexed caches can be isolated along with memory. \textit{Page colouring} takes the advantage of the overlapping bits between cache \textit{set} index and page index (for virtual-to-physical address translation). Pages can be assigned with different ``colours'' (i.e., the overlapping bits), and the colour decides which cache sets they are mapped to. Therefore, cache accesses are isolated along with memory access. Initially being used to improve system performance~\cite{bershad1994avoiding, kessler1992page, ye2014coloris}, page colouring has been re-purposed to build defenses, by isolating the cache that the untrusted code can directly access~\cite{kim2012stealthmem, zhou2016software}.

%\zl{software-based like page-coloring can't defend against our attacks, }

%Nonetheless, researchers have already noticed the threat and therefore designed cache isolation techniques with which cache attacks can all be defeated. 
%Cache isolation splits a cache set into multiple zones and each core can be associated with a zone. A core cannot touch the cache slots of other zones, even when other zones have empty slots. Cache isolation indeed wastes some cache and decreased cache utilization. However, cache slots for different users are isolated and non-volatile cache attacks can be difficult to launch. For example, prime probe based cache attacks can no longer be launched, because the prime step would not evict victim cache line belonging to another user.

\zztitle{Hardware Defenses}
Another perspective to defeat cache side channels is hardware-based isolation.
To provide better cache QoS, Intel has implemented a technique named  \textit{Cache Allocation Technology (CAT)}~\cite{cat}. It allocates different cache \textit{ways} to different COS (class of service). Each core is also associated with one or more COS. 
A core can access a cache way only when they share at least one common COS. Still, directly enforcing cache isolation with CAT is not straightforward, as the provided COS is limited to 4 or 16. CATalyst~\cite{liu2016catalyst} adjusted CAT to protect LLC by separating it into a secure and a non-secure partition, and forced the secure partition to be loaded with cache-pinned secure pages. On ARM, hardware mechanism like Cache Lockdown, can enable similar protection by pinning whole sets of the L1-I and L1-D caches~\cite{arm_lock}.

Hardware-based cache isolation and its extension are supposed to mitigate the attacks that bypass the software defenses. For example, by partitioning the page table and TLB with CATalyst, \textsc{Xlate}~\cite{van2018malicious} and \textsc{TLBleed}~\cite{gras2018translation} can be mitigated. To defend against directory-based attack~\cite{yan2019attack}, Yan \etal\ proposed SecDir to partition and isolate directories~\cite{yan2019secdir}.

In addition to CAT, another hardware feature of Intel, TSX has been used as defense~\cite{gruss2017strong}. Intel TSX introduces hardware transaction, in which case transactions would abort if it were interfered. By putting sensitive data and code in a transaction and pin it to a cache set, cache eviction triggered by adversaries will lead to abort. 

\zztitle{Temporal Isolation} Both hardware and software defenses aim at isolating resources spatially, which could not cover all resources. For example, small cache like L1~\cite{van2018malicious}, cannot be isolated due to not enough page coloring granularity. Ge \etal\ proposed to enforce temporal isolation with OS abstraction~\cite{ge2019time}, so the existing cache side-channels~\cite{yan2019attack,disselkoen2017prime+} can be mitigated when combining with spatial isolation techniques (hardware and software).
%In a word, cache side channels that rely on conflict of eviction sets can be well defeated by combining spatial isolation and temporal isolation, as shown in~\cite{ge2019time}.
However, the defense is only applicable to seL4 microkernel.

Table~\ref{tab:model_compare} summarizes the existing cache side-channels and how they are impacted under the existing defenses.

%For temporal defense, \cite{ge2019time} admits that they are powerless to stateless interconnect channels.

%Recent work found that CAT does not isolate directories for non-inclusive LLC, which results in cache attacks in directories. 
%Non-inclusive LLC does not allow a core to evict cache lines owned by another core in LLC, because cache lines in the private cache of another core may not exit in LLC. 
%However, as pointed by~\cite{yan2019attack}, the attacker is still able to evict another core's cache line by evict directories in LLC. 
%With partitioning deployed on directories~\cite{yan2019secdir}, directory based cache attacks can be defeated.

\subsection{Stateless Channels}
\label{subsec:stateless}

According to Ge \etal~\cite{ge2019time}, microarchitectural side-channels exploit the competition of hardware resources, which can be classified into two categories: \textit{microarchitectural state} and \textit{stateless interconnects}. The first category includes caches, TLBs, branch predictors, and DRAM row buffers, on which resource contention leads to the state changes observable to the adversary. The second category includes buses and on-chip networks. Though the concurrent access leads to a reduction of available bandwidth, no interference on the microarchitectural state should be observed. 

Exploiting stateless interconnects for attacks is not brand new. I/O bus contention~\cite{gray1993introducing, gray1994countermeasures,hu1992reducing,hu1992lattice} and memory bus lock~\cite{wu2014whispers} have been exploited to construct covert channels.
However, we found the research on this direction has been \textit{stalled}, probably due to that the exploited hardware features are outdated (e.g., VAX security kernel~\cite{gray1993introducing, gray1994countermeasures}). So far, no practical side-channel attacks through interconnect are known~\cite{ge2018survey}~\footnote{Wang \etal\ used a simulator to study the timing side-channel of on-chip network~\cite{wang2012efficient}, but the attack has not been demonstrated on the real CPU interconnect.}.
%Timing side-channel of on-chip network was studied with simulators. However, side channel on real mesh interconnect has never been studied before.
In this work, we aim to \textit{revive} this direction and demonstrate that stateless interconnects can be exploited for side-channel attacks.

%Wang \etal\ found the contention of the concurrent flows on on-chip-network may result in side channels~\cite{wang2012efficient}, but it was only demonstrated on a simulator. Moreover, it relies on that the cache is small, which is not true for the modern processors.

\ignore{
\zl{commented out for now, this might confuse the reviewers}
A recent work named CrossTalk~\cite{ragab_crosstalk_2021}  launched cross-core transient execution attacks, by sampling the staging buffer on the interconnect, which still exploits stateful micro-architectural information. \attack{} \textit{exploits} interconnect as a stepping-stone to attack  cache.
}

%Decades ago, Gray \etal~\cite{gray1993introducing, gray1994countermeasures} investigated the I/O bus contention between VMs managed by VAX security kernel, and leveraged it to construct covert channels. Memory bus lock can also be used to construct covert channel, as shown in~\cite{wu2014whispers}, though such lock cannot be used to construct side channel as normal apps rarely lock bus. 

%\attack{} exploits mesh interconnect to launch cache side-channel atta

%\zl{non-inclusive cache, but it still allows cache conflicts}

%\zl{Intel CAT and how it breaks existing attacks}

%\zl{talk about SecDir~\cite{yan2019secdir} to break ~\cite{yan2019attack}.}

%\zl{try to survey as much as you can about the cache defense, \eg~\cite{wang2016secdcp, yan2017secure, yan2019secdir}, and show we break all of them.}

\section{Attack Overview}
\label{sec:overview}

%All previous cache side-channel attacks assume the adversary shares either memory (Flush+Reload) or cache set (Prime+Probe) with the victim. 
%\zl{RA: ~\cite{wu2014whispers} has answered that}
In this section, we first introduce the threat model and compare it with the ones of prior related works. Then, we demonstrate why the leakage from mesh interconnect matters. Finally, we overview the steps of our attack \attack.

\subsection{Threat Model}
\label{subsec:threat}

%\zztitle{Attack Setting} 
We assume the attacker who intends to extract secret information, \eg, encryption keys, shares the same CPU with the victim, but resides in different cores from the victim. We envision \attack{} is effective in the cloud environment, when co-residency is achieved~\cite{varadarajan2015placement}.
We also assume \textit{all} the existing hardware and software defenses against the cache side-channel attacks are deployed and turned on, like page coloring and CATalyst.
%including the ones provided by the CPU vendors, like Intel CAT~\cite{cat}, and the ones proposed by the academia, like SecDir~\cite{yan2017secure}.
We target the latest high-end CPU, Intel Xeon Scalable CPUs, where core-to-core communication goes through mesh interconnects.
%and LLC is non-inclusive.

%The operating system is a multi-user one and different user accounts are given to the victim and attacker, and the standard protections like memory isolation and role-based access control are enforced. 
%We also assume exploiting a system/software level vulnerability is quite difficult, \eg, when all security patches are installed. As such, the attacker attempts to achieve her goal by launching cache side-channel attacks when the victim's program is running. The above assumptions are the same as previous cache side-channel attacks~\cite{liu2015last,yan2019attack}.

%provides strong enough isolation between different users, like the Linux file system. 
%\zl{this setting is too specific}
%The victim runs a program that processes sensitive data. For example, the victim program may be a RSA decryption or digital signature program, which uses a private key that is assumed to be confidential to the attacker.
%Therefore, the attacker is not able to directly access victim's sensitive data. The attacker aims at stealing the victim's sensitive data, but does not exploit vulnerabilities of the operating system.

\begin{figure}[ht]
  \centering
  \includegraphics[width=0.45\textwidth]{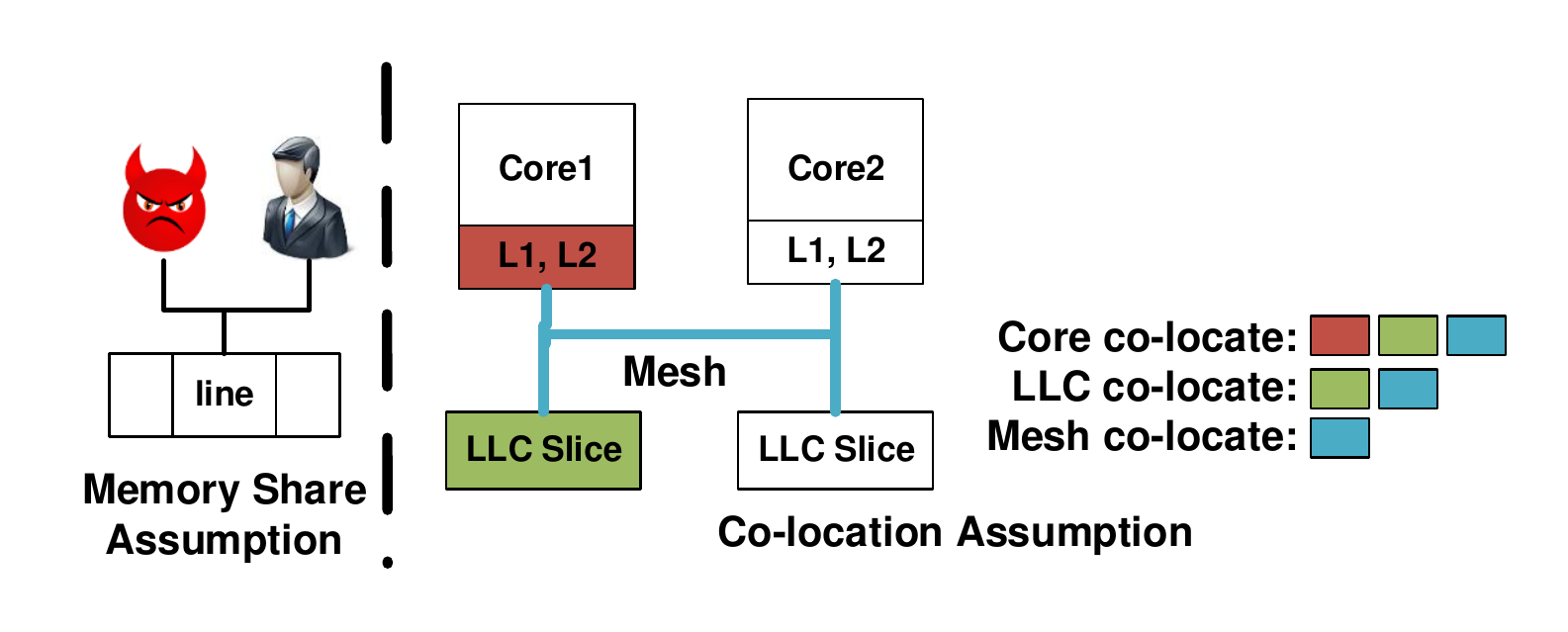}
  \caption{Comparison of the settings between different cache side-channel attacks.
  %\zl{appendix}
  %\zl{can you redraw it? it still doesn't look good}
  }\label{fig:assumption}
\end{figure}
%\zztitle{CPU Co-locate Assumption} 

Here we compare our setting to the existing cache attacks. The strongest assumption made by the prior works is the sharing of memory addresses (shown in Figure~\ref{fig:assumption} left), like \textsc{Flush+Reload}. 
However, memory sharing can be turned off for the critical data/code. A weaker assumption is that cache sets are shared (shown in Figure~\ref{fig:assumption} right), so the attacker can evict cache lines of the victim (or vice versa), like \textsc{Prime+Probe}. Under this assumption, the attacker either shares in-core private L1/L2 cache with the victim~\cite{osvik2006cache}, or out-of-core LLC (either inclusive LLC~\cite{irazoqui2015s} or non-inclusive LLC~\cite{yan2019attack}). However, the assumption does not hold when spatial or temporal resource partition is enforced, as discussed in Section~\ref{subsec:isolation}.
%\zl{appendix, Section~\ref{subsec:isolation} has demonstrated all attacks can be defended.}\todo{I think It's OK to put it here.}

\zztitle{Stateless Cache Side-channel Attack} As described in Section~\ref{subsec:stateless}, the existing attacks all lead to microarchitectural state changes with resource contention, and we call them \textit{stateful} attacks. 
\textbf{Can we launch cache side-channel attacks without touching the cache state of a victim program?} It seems impossible at first sight, but the mesh interconnects render \textit{stateless} attacks possible. 
\attack{} only requires that the attacker and victim share routes in mesh interconnect, which, as we show later, is easy to achieve. This setting entails the sharing of CPU, an assumption unlikely to be invalidated by the defender. \attack{} is expected to \textit{bypass page coloring, hardware isolation and also temporal isolation}. 
\attack does not assume eviction set conflict between victim and attacker, so page coloring does not help. Similarly, \attack{} can bypass hardware defenses like CAT, and part of our evaluation experiment was done with CAT on. \attack{} can also bypass TSX protection, as it does not touch (evict or reload) any victim cache lines. 
Though the time protection of~\cite{ge2019time} is very effective against the existing stateful cache side-channel attacks, the authors admit that they are \textit{``powerless''} against stateless channels, since there is no ``appropriate hardware support" to partition interconnect (buses).

\subsection{What Leaks from Mesh Interconnect?} 
\label{subsec:rationale}

%, as shown by Figure~\ref{fig:reasoning}.

\ignore{
\zl{we are over page limit}
\begin{figure}[ht]
  \centering
  \includegraphics[width=0.45\textwidth]{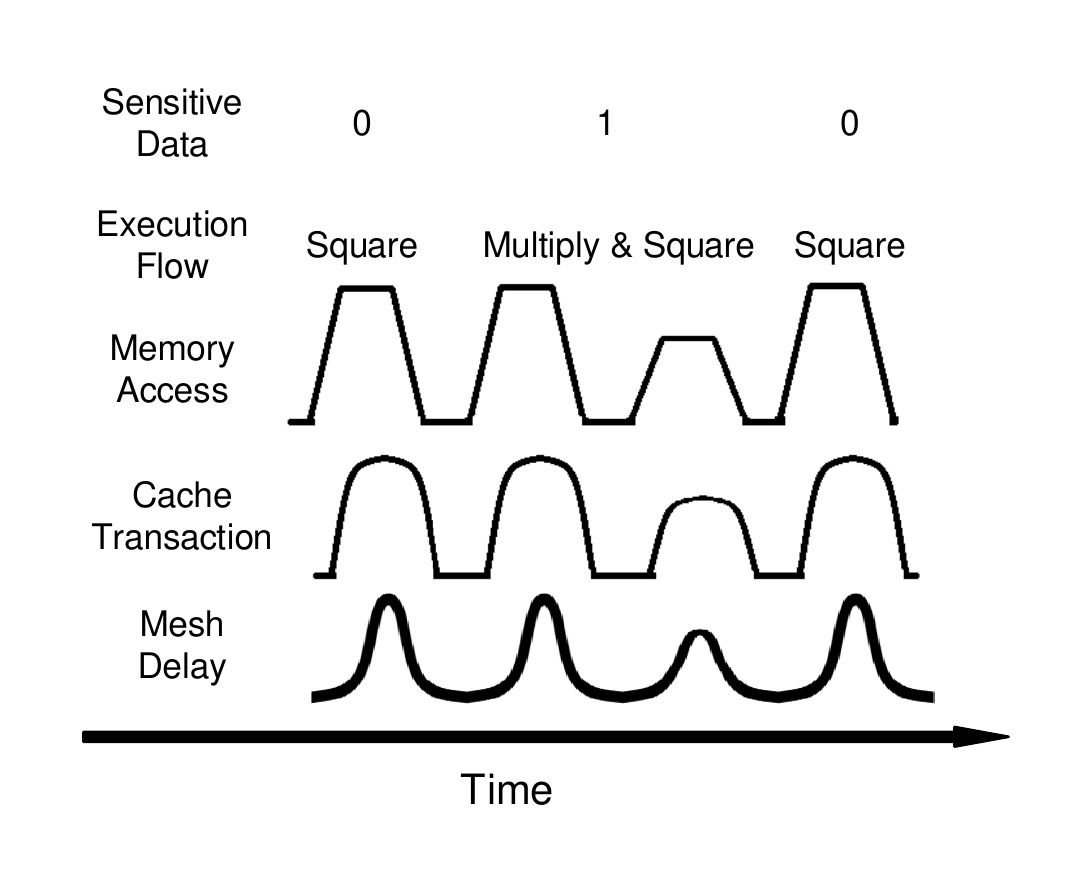}
  \caption{The relation between sensitive data, execution flow, memory access, cache transactions and mesh delay, when RSA is running.
  %\zl{this graph looks random,  can you map the RSA leakage in your example to the mesh traffic on the routes?}
  }\label{fig:reasoning}
\end{figure}
}

%. Thus, the execution flow sequence is directly corresponded with the private key. As a result, when the execution flow acquired by the attacker, the private key can be recovered bit by bit, 

%why mesh traffic matters, reference to the section of cache access, plenty of places

%Though the sensitive data consumed by a victim program could be well protected, the memory and cache access patterns could leak information about the sensitive data, especially when the data decides the program execution flow. 
It has been demonstrated by the existing works that the data/code access patterns leak sensitive information of a program. Take the RSA fast modular exponentiation algorithm as an example (as shown in Appendix~\ref{sec:mod_exp}), it iterates every bit of the exponent $d$ of a private key. If the bit is 0, the program executes the square and mod functions. When it is 1, the program will execute a round of multiply and mod function in addition to the square and mod function. Thus, bit 0 and 1 yield different memory access patterns. Memory access also triggers cache transactions, and by monitoring the cache occupancy, victim's memory access patterns can be reversed.

%and different execution time. Associated with memory access, cache transactions also happen. The previous non-volatile cache attacks evict the cache lines of these instructions, and probe them later to decode the execution flow, till revealing the key bits. However, under cache isolation, an attacker cannot evict the lines in the cache (L1/L2/LLC) associated with another program

%. Is the leakage blocked? \textit{Not so.}
The existing defenses are able to block the contention by the attacker on cache and other associated units like MMU, so the victim's cache state cannot be monitored. However, the bandwidth on mesh interconnect cannot be segregated between users.
%As described in Section~\ref{subsec:cache}, cache access involves transactions over mesh interconnect, which is shared \textit{across cores}, and \textit{none of the defenses segregating the mesh traffic by applications}. 
If an attacker program can intentionally inject mesh traffic at a fixed rate to contend with the traffic generating from/to a victim program,  when the victim's traffic volume increases, \eg, loading more data from memory, the attacker should sense the delay increase of its own requests, given that the maximum bandwidth of a mesh link is fixed. When the victim's traffic volume decreases, the delay sensed by the attacker should also be reduced. Hence, the memory access patterns can be inferred from the mesh delays. %Figure~\ref{fig:reasoning} illustrates such relation.

\subsection{Challenges}
%Mesh Traffic Estimation.
%\zl{to be edited}
%layout reverse-engineering
Though creating the contention-based side-channel on mesh interconnect seems easy, there are multiple challenges we need to address. 1) The attacker needs to know where the victim program is located on the mesh layout, and where the victim traffic flows from and to. But the mapping between the high-layer program primitives, like processes and threads, to the low-layer hardware units, like mesh tiles, is obscured. In fact, Intel even keeps the CPU layout, \eg, which tile has which core, as proprietary information. Besides, the relation between cache transactions and mesh traffic is unknown. 
2) There is no ISA support for the attacker to generate \textit{direct} mesh traffic between two stops. The attacker can only access her own memory to indirectly produce mesh traffic. Though Intel provides a tool, PMON~\cite{pmu_manual}, for users to profile mesh traffic, the statistics are aggregated over a long period of time. Besides, PMON can only be used by root users. We want the attacker to be able to sample the mesh traffic at high frequency as a \textit{non-root user}. 
3) Comparing with the stateful cache side-channels, where the operand address of an instruction can be evicted and probed individually, \attack{} only has visibility to traffic from/to a core, which is more coarse-grained. 
We will address the above challenges with novel techniques in reverse engineering, probe design, and decoding.

%but only root users can access those APIs in Linux. Besides, PMU does not provide real-time access to root users. Root users can only get statistic about mesh traffic in a long period of time.

%destination

%coarse-grained decoding
%\zl{PMU change to PMON}

%To overcome the challenge, we propose that volatile cache attackers can measure the delay of a series of cache transactions and infer victim's private data according to delay sequence. As for the reason, the delay of a cache transaction is highly correlated with the co-exist traffics along the path. Cache transactions involves cache line transportation along a path over mesh. The delay would be higher if the path for a transaction also carries other traffics at the same time. The attacker can elaborately design cache transactions in a way that her traffics encounter victim's traffics with high chance, making the collected delay sequence correlated with victim's cache transactions.
%\zl{this part is quite unclear to me}

%\zl{I'll work on this section, after you finish. You should focus on the mesh leak, which is the most interesting part.}

\subsection{Attack Steps}
\label{subsec:steps}

%\zl{to be edited}
\attack{} consists of three steps summarized below. Step 1 is done before attacking the victim.

\zztitle{Step 1: Layout Reverse-engineering (Section~\ref{sec:reverse})} Before launching \attack, the attacker could build the mapping between a core/CHA ID to its geometric location on the CPU layout, to help the later stage. We use core/CHA IDs to fill the layout map because they are bounded with applications, and the tools to obtain the IDs are accessible to non-root users. 
Since the layout should be the same for all CPUs of the same model and stepping level, this only has to be done once. 

\zztitle{Step 2: Probe \& Measurement (Section~\ref{sec:probe} and~\ref{subsec:mesh_impact})} 
When launching attacks, the attacker runs an application on the target machine, sharing CPU with the victim. The attacker could first identify the geometric location of the cores hosting the victim application, 
by querying the core ID associated with the victim process and locate the tile on the reverse-engineered layout. The information is leveraged to identify the paths that contend with the victim's mesh traffic at high probabilities. 
If the victim process is invisible to the attacker, \eg, on the virtualized platforms, the attacker can select a random path. Our evaluation shows the information leakage from the random path contention is still prominent.
%\zl{please check;Done}
To direct mesh packets over the selected paths, the attacker constructs an eviction set, and probes the memory addresses within it. The probe is issued repeatedly and the interval consumed by each probe is logged.
To notice is that \textbf{the eviction set does not conflict with the victim in the cache}, hence the existing defenses can be evaded.

%, when cache isolation is enforced. Therefore, accessing the set does not interfere the cache status of victim.

%potentially carries most of the victim's traffics. Then, she can choose a sub-path within the identified key path and tries to produce cache transactions that incurs traffics along the sub-path. 
%At the same time, the attack record the timing of the completion of each transaction she made.

\zztitle{Step 3: Secret Inference (Section~\ref{sec:evaluation})} 
%The last step is sensitive data inference. 
After the last step, the attacker obtains a sequence of intervals, and the secret underlying the sequence is to be decoded. This step is application-specific, as different victim programs produce different delay sequences. 
This step can be done at the attacker's own machine.
In particular, we use RSA to showcase how to decode a delay sequence.

%To infer the sensitive data according to the collected delay sequence, the attacker should learn the mapping between the delay sequence pattern and private data on her own machine. Then, she can map the delay sequence to the private data.

%\zl{I'll work on it after finishing the later sections.}
\section{Reverse Engineering the CPU Layout}
\label{sec:reverse}

%For better performance, the attacker could firstly make clear which tile has victim's data/code and which route the victim traffic would go through, so that attack traffic can be crafted to congest specific routes of the mesh interconnect. 
%To launch volatile cache attacks, an attacker must co-locate her mesh traffic with the victim's. To do such a co-location, the attacker must know where the victim's core is. 
%Besides, the attacker needs to generate traffics precisely on a path over mesh, 
%As a prerequisite, the layout of CPU, in particular, the mapping between tiles and cores, has to be known ahead by the attacker. This is however kept as proprietary information by the CPU vendors. 
In this section, we propose a new method to reverse engineer the CPU layout to help the later attack stage. 
To notice, this step requires root privilege to execute some profiling instructions, but during the actual attack, the root privilege is unnecessary. This step only needs to be carried out once for a targeted CPU model (and stepping level). We demonstrate the method on two Intel Xeon Scalable CPUs, 8260 and 8175, and the two steps are elaborated below. Due to the space limit, we leave the layout of 8175 in the appendix, as 8260 is the CPU for our attack evaluation.

%\zl{need to talk about the primitives and justify whey they are useful}
%\zl{what part are documented? and what part are inferred by us? we need a clear line}

\subsection{Identifying the Enabled Tiles}
\label{subsec:tileid}

There are three types of CPU dies, named LCC, HCC, XCC (for low, high, and extreme core counts), for Intel Xeon family, with 10, 18 or 28 cores in a die respectively~\cite{xcc_wiki}. However, when a CPU is shipped to the customers, Intel might intentionally disable some cores. For example, the Xeon Scalable 8175 CPU has 24 active cores, because Intel disabled 4 cores. Without knowing which cores are enabled, the attacker cannot find the best route to congest. 

Here, we exploit the hardware feature of Intel CPUs to reveal such information. According to Intel's document~\cite{pmu_manual}, a user can query the \texttt{CAPID6} register to learn the ID of the tile whose CHA is disabled.
%which CHA is disabled. 
When this is the case, the whole tile including the core and LLC inside are also disabled.
Take Xeon Scalable 8175 as an example. Its \texttt{CAPID6} contains 28 bits to indicate the status of all tiles, and a CHA is disabled if its associated bit is 0.
By reading all bits of \texttt{CAPID6}, we found  
%28 bits for CPUs with HCC configuration \ie CPUs with 19-28 cores. 
bit 1, 4, 24, 27 are set to 0. 
%But the question here is where these tiles are located in the CPU. 
Then we leverage a previous research~\cite{mc_ppt}, which maps tiles to the CPU layout, and mark the tiles as disabled based on the \texttt{CAPID6} bits. 
%meaning that these four tiles are disabled (a tile includes a core, a CHA and its LLC and SF). 
%According to , the tiles are horizontally displaced in the die, 
%resulting in the layout shown by 
Table~\ref{tab:possible_pos} in Appendix~\ref{appendix:xs8175} shows what has been inferred on Xeon Scalable 8175.
For Xeon Scalable 8260, though it has the same number of cores, the disabled tiles are different, (the IDs are 2, 3, 21 and 27). 
%\zl{[RB] doesn't think the list of disabled tiles depends on the model, page 15 of~\cite{mc_ppt}. I added a restriction that must be same model and stepping level.}
%According to the tests of 62 Xeon 8160 CPUs~\cite{mc_ppt}, the layout is identical for every CPU of the same model (\eg 8175 or 8260), therefore, we conclude the attacker does not need to execute this step again if one CPU of the targeted model has been reverse engineered.

%Dr. McCalpin tested over  and also confirmed the layout is fixed for the same model CPUs.

\subsection{Mapping CHAs and Cores to Tiles}
\label{subsec:numbering}

As described in Appendix~\ref{sec:access}, a requesting core talks to the CHA of a target core to initiate a cache read transaction. To direct the transaction between two tiles for mesh traffic, the attacker needs to know the ID of the requesting core and the ID of the target CHA. Yet, the relation between the core/CHA IDs and the tile IDs is not documented anywhere. We infer such information in this step.

According to~\cite{mc_ppt}, CHAs are sequentially numbered along with tiles, but when a tile is disabled, the CHA ID is skipped. As such, CHA is numbered from 0-23 for Xeon Scalable 8175, and tile \#2 has CHA \#1 because tile \#1 is disabled. Table~\ref{tab:cha_cpu_pos} in Appendix~\ref{appendix:xs8175} shows the mapping of CHAs to tiles of Xeon Scalable 8175, and the first number of each pair is CHA ID. 

%To make sure that the ID of CHAs and cores does not exceeds the total enabled number of tiles, 

Regarding cores, the task becomes non-trivial, as they are \textit{not sequentially numbered}. 
McCalpin proposed a method to infer how the cores are aligned by reading ``mesh traffic counters''~\cite{mc_ppt}. However, the author also admitted the result needs to be disambiguated~\cite{core_l3_numbering}. 
To improve the accuracy of the inferred layout, we propose a new method. Specifically, we bind a thread to a core (by setting its affinity~\cite{thread_affinity} to the core ID), and use it to access 2GB memory. We monitor the Intel performance counter \texttt{LCORE\_PMA GV}~\cite{pmu_manual} and found the CHA yields the highest value when it co-locates with the core in the same tile. \texttt{LCORE\_PMA GV}\footnote{It is the abbreviation of Core Power Management Agent Global system state Value, according to ~\cite{pmu_manual} } indicates the global power states of the core. When the increase of \texttt{LCORE\_PMA GV} is observed, we assign the core ID to the tile. We repeat the process for every core, and the layout can be reconstructed. The second number of each pair shown in Table~\ref{tab:cha_cpu_pos} in Appendix~\ref{appendix:xs8175} is the core ID, for Xeon Scalable 8175. In Table~\ref{tab:cha_cpu_pos_2}, we show the inferred layout for Xeon Scalable 8260, which is different from 8175.

%to access a huge page memory chunk and found that the 
%Therefore, we learn that the core is co-locate with the CHA in a tile. 
%By repeat the test on each core, we constructed the layout of cores, as shown by 

%\subsection{Reverse on the Latest CPUs}
%Besides Skylake-SP CPUs, we also repeat the revere engineering on the latest generation of Intel Xeon Scalable CPUs \ie Xeon Cascade Lake Scalable CPUs, 
%and found the numbering scheme is the same with Skylake-SP. Table~\ref{tab:cha_cpu_pos_2} shows the results.

\begin{table}
  \centering
  \begin{tabular}{|c|c|c|c|c|c|}
     \hline
     UPI  & PCIE & PCIE  & RLINK & UPI2 & PCIE \\ \hline
     0, 0    & 2, 16 & 7, 19  & 12, 3 & 17, 16 & 21, 17 \\ \hline
     IMC0 & 3, 18 & 8, 2 & 13, 15 & 18, 10 & IMC1 \\ \hline
     1,12    & 4, 1 & 9, 14 & 14, 9 & \cellcolor[gray]{0.8} & 22, 11 \\ \hline
     \cellcolor[gray]{0.8}    & 5, 13 & 10, 8 & 15, 21 & 19, 22 & 23, 23 \\ \hline
     \cellcolor[gray]{0.8}    & 6, 7 & 11, 20 & 16, 4 & 20, 5 & \cellcolor[gray]{0.8}  \\ \hline
   \end{tabular}
  \caption{Layout of Xeon Scalable 8260 CPU, with the same setting as Table~\ref{tab:cha_cpu_pos}. }\label{tab:cha_cpu_pos_2}
\end{table}

%As we can see from Table, the mapping between CHA ID and core ID is the same for the two CPUs of different generation. However, Intel disabled different physical tiles, though they have the same number of cores (24) and the same die configuration (XCC). Therefore, when launching attacks, the attacker can not assume CPUs with same number of cores share the same layout. 
%She must do the preparation on a CPU with exactly the same model as the target machine. 

\section{The Probe Design}
\label{sec:probe}

After the layout of the target CPU is inferred, the attacker uses a probe to contend with the mesh traffic generated from/to the victim's core, and measure the probe latency to infer the victim's activities. 
In this section, we describe the probe design. 

%In Table~\ref{tab:symbols} we summarize the symbols used in this section.
\ignore{
\begin{table}[t]
    \centering
    \begin{tabular}{c|c}
         Symbol  &  Description   \\ \midrule[1.5pt]
         $r$ & The requester   \\ \hline
         $t$ & The target  \\ \hline
    \end{tabular}
    \caption{Symbols used in Section~\ref{sec:probe}.}
    \label{tab:symbols}
\end{table}
}

\subsection{Characterization of Mesh Traffic}
\label{subsec:path}

%\zl{a table about symbols}

%The attacker aims at learn the victim's memory access pattern through mesh traffic delay. Wanting the delay interfered by the victim's mesh traffic, she must design probe traffic such that her traffic co-locate with the victim's.
%To congest the route that victim's cache transaction passes by, the path selected by the attacker should receive a prominent portion of victim's traffic.

How to select the paths that contend with the victim's mesh traffic is a non-trivial problem, since different types of traffic are involved and each tile issues/receives traffic from different directions. Below we introduce the types of mesh traffic (termed \textbf{T1-T7}), summarized from the existing documents~\cite{mesh_textbook}. Then, we describe how traffic is distributed and the strategy of path selection. We use $r$ and $t$ to refer to the entity issuing the request and the target. Hence, the requesting core is termed $Core_r$, and the co-located L2, LLC slice and CHA are termed $L2_r$, $LLC_r$ and $CHA_r$. For the target, the terms are $Core_t$, $L2_t$, $LLC_t$ and $CHA_t$. We focus on the movement of cache lines or memory blocks, which take a much larger size than the mesh messages. The typical flow of cache access is described in Appendix~\ref{sec:access}.

\begin{itemize}
    \item \textbf{T1: $Core_r$ to $LLC_t$.} When $Core_r$ encounters L2 cache miss, it will compute the LLC slice ID and send messages to $CHA_t$ (same or different tile with $Core_r$) co-located with $LLC_t$, asking if the cache line is presented. Also, when $L2_r$ is about to evict a line to $LLC_t$, \eg, when the L2 cache set to be inserted is full, the memory sub-system of the core will pass the line
    %\footnote{Victim cache line is the cache line selected to be evicted, not the cache line belongs to the assumed victim in the attack.} 
    from $L2_r$ to $LLC_t$ (same or different tile with $L2_r$).
    %Both cases incur traffics from the core to the LLC slice.
    
    \item \textbf{T2: $LLC_t$ to $Core_r$.} Following T1, if the cache line is in $LLC_t$, $LLC_t$ will send the cache line to $Core_r$.
    
    \item \textbf{T3: IMC to $Core_r$.} Alternatively, if the cache line is not present in $LLC_t$, $CHA_t$ will ask the IMC to fetch the line from memory and send it to $Core_r$. To be noticed is that the line is directly sent to $Core_r$, when LLC is non-inclusive.
    
    \item \textbf{T4: $LLC_t$ to IMC.} When $LLC_t$ is full, to accept new cache line insertion, it will evict the least recently used cache line to the IMC.
    
    \item \textbf{T5: $Core_r$ to $Core_t$.} $Core_r$ can access a cache line in the private cache of $Core_t$ when the they share memory. $L2_t$ will pass the cache line to $Core_r$ through mesh.
    
    \item \textbf{T6: $LLC_t$ to I/O Unit.} Intel CPUs allow I/O devices to directly access LLC and bypass memory for better performance, under Data Direct I/O (DDIO)~\cite{gilbert2012intel}. In this case, cache lines will be passed between a PCIe stop (stop inside a PCIe tile) and $LLC_t$.
    
    \item \textbf{T7: Other traffic.} It characterizes the mesh traffic undocumented by Intel, which is expected to have a smaller volume than T1-T6.
\end{itemize}

%\zl{RA: formalism is too excessive}

%\zl{what types of victim applications fit your assumptions.}
While different applications exhibit different memory access patterns, we found the applications we are interested in, \eg, RSA, \textit{distribute most of the mesh traffic through T1-T3}.
%\zl{we need a better characterization of applications.} 
%We profile RSA on processor xxx using Intel PMU~\cite{pmu_manual} and the statistics of mesh traffic is shown in Figure~\ref{}. \zl{result with Intel profiler.} 
Below we use RSA as an example. Since it repeatedly accesses the same region of data/code (see Appendix~\ref{appendix:rsa}, which has a while loop), most of the accesses can be served by L1/L2/LLC, scaling down the share of T4 and T6. The implementation of RSA does not use multi-threads, so T5 is also small.
%In addition to CPU profiling, we also estimate the traffic distribution using the parameters of the application, \ie, RSA, and the Intel CPU, and show the result in Figure~\ref{}.
%\zl{result with the theoretic analysis.}
On the other hand, we acknowledge there are applications with different patterns, \eg, more traffic on T4 - T6,  and we discuss this case in Section~\ref{sec:discussion}.

%The attacker should firstly identify which paths carry most of the victim core's mesh traffic. After our analysis, the first three kinds of mesh traffic may dominate. \zl{can we have an experiment?}

%The reasons are as follow: 1) Traffic from IMC to LLC could be small, as Xeon processors have very large overall cache size, especially the shared LLC. Most memory access can be served by the three levels of cache. 
%For memory intensive applications, the traffic between IMC and LLC could be large. But in this work, we consider general apps only.
%2) Traffic between cores is totally attributed to cross-core cache line share, like mutex. The ratio of shared memory access is low for normal applications, meaning not much cross-core traffic on mesh. 
%For some cross-core sharing intensive application, the fifth kind of traffic may dominate. 

That T1-T3 shares the majority of mesh traffic is an ideal condition for \attack{}, because they are directly related to the core occupied by the victim application. Therefore, we can select the attack path based on where the victim core is located and the direction of mesh traffic from/to the core. To make \attack{} stealthy, we aim to use the least number of cores to construct the attack paths. A key question here is \textit{how the traffic from/to victim's core is distributed in the 4 directions.} We attempt to answer this question based on analyzing the geometry and routing pattern of the evaluated CPU. In Section~\ref{subsec:location}, we use a profiling experiment to support our theoretical analysis.

\begin{figure}[ht]
  \centering
  \includegraphics[width=0.45\textwidth]{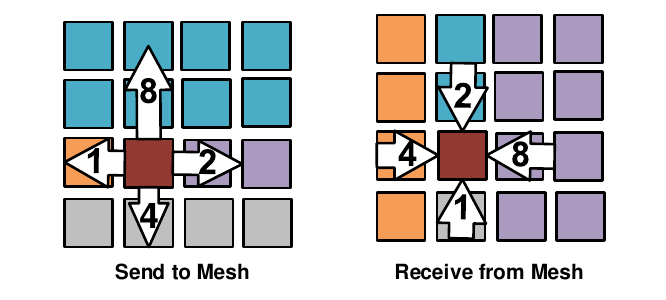}
  \caption{Mesh traffic to and from a core. The number inside the arrow show the number of tiles that may receive/send traffic.}
  \label{fig:path}
\end{figure}

For T1, all LLC slices should have a similar probability to be visited, because the LLC hashing function designed by Intel strives to achieve this goal~\cite{horro2019effect}.
%\zl{can you check the citation?}
Hence, \textit{the traffic volume going out from a core in one direction is proportional to the number of tiles reachable on that direction}. According to YX routing, a mesh packet first goes vertically to the row of the target tile. For one vertical direction (north/south), assuming the number of tiles per row is $T$, the number of rows is $R$ and the number of disabled tiles is $D$,
%\zl{[RB] If it is the number of disabled cores on the die, the formulas are correct.  If it is the number of disabled cores in the relevant area, its value is overloaded.} 
the estimated traffic volume $S$ from the victim core will be $S=T \times R-D$. Then the mesh packet goes horizontally at the same row of the target. For one horizontal row (east/west), the traffic volume from a core will be $S = C-D'$, where $C$ and $D'$ are the numbers of columns and disabled cores on the row. Figure~\ref{fig:path} left shows an example. 

\ignore{
%After our analysis, the vertical path that is farther away from the die boarder carries most of the outgoing traffic of a core, while the horizontal path that is farther away from the die boarder carries most of the incoming traffic. 
the two paths are deemed as the key path for volatile cache attacks. 
Specifically, the traffic between a core and a LLC slice (\ie the first two kinds) is evenly distributed across all the slices in the die, thanks to the LLC slice hash. 
However, the traffic is not evenly distributed, as shown in the left of Figure~\ref{fig:path}. 
For traffic sending out from a core (\ie L2 cache miss messages and evicted L2 lines), more than half of the traffic will be sent out through the vertical path that is farther away from the border.
%\zl{we need an experiment to prove your claim}
Specifically, mesh traffic is always firstly vertically sent to the row of the destination LLC slice, indicating that more than half LLC slices (\ie LLC slices above the requesting core) should be reached through vertical path first. 
What makes the traffic even more unbalanced is the position of the core. When the core is at the lower (upper) part of the die, there are more LLC slices above the core. Therefore, the upper (lower) part LLC slices predominate and the traffic to them all goes through the north (south) direction first.
}

For T2, the distribution of traffic volume is reversed, as the core becomes the receiver. The packet from every tile east/west to the core has to move vertically firstly to the row where the core resides, and then horizontally till reaching the core, so the traffic volume to the core from east/west is $S=T \times C-D$ ($T$ being the number of tiles per column now). The remaining packet flows to the core only vertically, so the volume becomes $S=C-D'$. Figure ~\ref{fig:path} right shows an example.

\ignore{
Similarly, the traffic from LLC slices to the requesting core is also unbalanced, as shown by the right of Figure~\ref{fig:path}. The vast majority of traffic would be received from the horizontal direction. For a core in the left (right) half of the die, more traffic will be received from right (left), \ie from the direction that is farther from the die boarder.
}

%\zl{to be edited}
For T3, our analysis is different. We assume that the LLC miss rate is $x$. The probability $p$ of a LLC slice being responsible to send a cache line is $(1-x)/n$ where $n$ is the number of cores, as the LLC hashing algorithm gives each slice an equal chance to serve a L2 miss. 
Then, the pattern of T3 depends on $n$ and $x$. For Xeon Scalable CPUs, $n=24$. When $x$ is around 4\%, $p$ will be a similar value, so the traffic from it would resemble T2.
If $x << 4\%$, the traffic from IMC could be ignored. 
When $x$ is significantly more than 4\%, the traffic volume $S$ should be increased proportionally to $x/4\%$, though the ratio between directions is the same as T2.

When the attacker program is able to learn the core ID, the optimal paths can be selected to increase the chances of bandwidth contention. Appendix~\ref{appendix:access} describes the details. However, in the virtualized environment, the core ID cannot be learnt. Yet, our analysis in Section~\ref{subsec:location} shows the impact of lacking such knowledge is moderate. To simulate this situation, during evaluation, we assign a random core ID is to the attacker and make the victim core ID oblivious to the attacker.

%\zl{we need an equation to reflect how latency is composed, including the mesh latency, the cache latency}

%\zl{We need an example here showing how the attack path is selected.}

\subsection{Probe based on Cache Eviction}
\label{subsec:traffic}

\begin{figure}[h]
  \centering
  \includegraphics[width=0.45\textwidth]{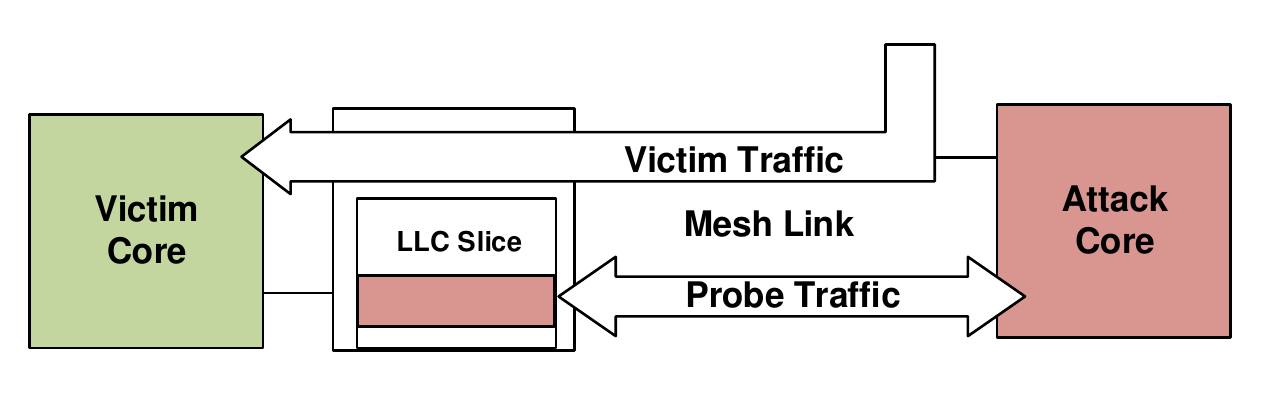}
  \caption{The probe designed by the attacker.%\zl{change attack to attacker} 
  %\zl{can be nicer... at least change the font?}
  }\label{fig:probe}
\end{figure}

%In this step, we show how to generate traffic to the designated CHA. 
%In this step, we describe how the probe is constructed.  
Different from routing a packet on the Internet, in mesh interconnect, a program \textit{cannot} explicitly sends traffic to the destination, because the CPU chip determines the route. To address this challenge, we adapt the existing methods for \textit{constructing evictions sets}~\cite{liu2015last, yan2019attack}. An eviction set is a set of memory addresses that are mapped to a cache set and able to evict all lines of the cache set. Because the cache set is set-associative, at least $w$ addresses are needed for an eviction set if a set has $w$ ways. 
Previous attacks, \eg, \textsc{Prime+Probe}, use an eviction set to evict lines of the private caches in the victim core. 
Though \attack{} uses eviction set, our goal is \textit{not} to evict victim addresses. To the contrary, \textbf{\attack{} evicts lines of its \textit{own} private L2 cache}, in order to generate mesh traffic of T1-T3 flowing to a designated LLC slice on the selected path. As such, \attack{} stays out of the protection realm of any existing defenses (see Section~\ref{subsec:isolation}). In Figure~\ref{fig:probe}, we illustrate the concept of our probe.

%mention the goal is different, not to conflict with victim's cache

%\subsection{L2 Eviction based Probe}
%To precisely generate mesh traffic between a core and a LLC slice, the attacker can cause L2 misses that all points to the same LLC slice. 

\zztitle{Constructing Eviction Set}
We use the approach described as follows to construct an eviction set. Firstly, the attacker prepares a set of memory addresses (denoted as $EV$)  that are mapped to one L2 cache set. 
The number of addresses (denoted as $n$) in $EV$ is set to be larger than the number of ways (denoted as $w$) a L2 set has (\ie, $n > w$), therefore when requesting addresses of $EV$, L2 cache misses always happen after $w$ requests. From $w+1$ to $n$ requests, each time a line is evicted from L2 to LLC, and a new line from memory will be inserted to L2. After that, when requesting $EV$ again, $n$ lines will be evicted from L2 to LLC, and $n$ lines will be passed from LLC to L2 in return, resulting in stable bi-directional mesh traffic on the attack path. 

\attack{} needs to force all L2 misses to be served by one LLC slice. Thus, $EV$ is not only mapped to a set of L2, but also a set of a LLC slice. 
%We first consider $EV$ for a LLC slice. 
To find addresses for such $EV$, we use the two routines proposed in ~\cite{yan2019attack}, \texttt{check\_conflict} and \texttt{find\_EV}, which are designed for non-inclusive LLC. In essence, \texttt{check\_conflict} tries to test if removing an address of a set makes cache conflict disappear. \texttt{find\_EV} tries to utilize \texttt{check\_conflict} to filter out a set of addresses that are all mapped to the same LLC slice. In Appendix~\ref{appendix:routines}, we describe them in detail.
To also force $EV$ mapping to only one L2 set, we split the $EV$ into $EV_0$ and $EV_1$ of equal size, and set the 16th bit different (0 and 1). 
As shown in Figure~\ref{fig:address}, bits 15:6 of a memory address points to a L2 set, while bits 16:6 points to a LLC slice set. As such, $EV_0$, $EV_1$, and their union $EV$ are associated with a single L2 cache set. 

%The attacker can get two equal large LLC eviction sets ($U_1$ and $U_2$) that have the same 15 to 6 bits addresses. In other words, the addresses in $U_1$ and $U_2$ has the same 15 to 6 bits, but the 16 bit is set to zero and one respectively. The union ($U$) of $U_1$ and $U_2$ is used to generate probe traffic. 

%The \texttt{check\_conflict} algorithm in~\cite{yan2019attack} combined with the eviction set algorithm in~\cite{liu2015last} helps the attacker get an eviction set whose addresses are all mapped to the same LLC slice and set, as showed in~\cite{yan2019attack}. 
%To be noticed is that a L2 set correlates with two LLC sets, as the set index of LLC uses one more bit (bit 16) than L2 set index.

%Specifically, the L2 set would be firstly fulfilled by the first $w$ lines with L2 and LLC misses. From the $w+1$th access on, until the $n$th access, a line will be evicted from L2 to LLC because of full L2 set and a new line will be inserted from memory, during which the L2 set remain full. 
%After that, every access to $U$ causes a line evicted from L2 and a line supplied from LLC to L2, resulting in stable bi-directional mesh traffic. 

Even when all cache lines fall into one LLC set, the attacker cannot decide which LLC slice serves the $Core_r$ yet. Though as shown in Figure~\ref{fig:address}, there is a hash function mapping a memory address to a LLC slice ID, the hash function is not disclosed, and the difficulty of reverse engineering is tremendous~\cite{yan2019attack}. Instead of recovering the hash function, \cite{yan2019attack} suggests testing an $EV$ (multiple $EV$ have been constructed as candidates) to see if it co-locates with $Core_r$. We let the attacker enumerate every core on her own CPU, and measure its access time to an $EV$. When the serving LLC slice is local to the core, the access latency is the lowest, and we link $EV$ to the LLC slice ID.

%The attacker also wants the L2 misses to be served by her appointed LLC. ~\cite{yan2019attack} showed how to learn if an address belongs to the LLC slice that co-locate with the core in the same tile, which allows the attacker to know which LLC slice $U$ belongs to, according to the reverse engineered tile layout (\ie Table~\ref{tab:cha_cpu_pos}). The attacker can generate different $U$s and choose the one belonging to her wanted LLC slice.

Finally, the size of $EV$ ($n$) has to be tuned carefully by the attacker. Not only $n$ should be larger than $w$ to force L2 misses, it should also avoid being too large to overflow LLC and L2 together. Otherwise, a line will be evicted from LLC to memory (T4), a request that takes much longer time to respond than T1-T3, reducing the probing frequency. As the Xeon Scalable Professors use 11-way LLC set and 16-way L2 set~\cite{xeon}, we varied $n$ of $EV$ from 18 to 38 (2*11+16), and found when $n$ equals to \textbf{24} ($EV_0$ and $EV_1$ each has 12 addresses), the mesh traffic generated within an interval is the highest.

\zztitle{Measuring Mesh Traffic} 
%timestamping
When the victim application is running, the attacker sequentially visits every address within $EV$ immediately after receiving the response to the prior request, and records the timestamp of each request (\eg, using the instruction \texttt{RDTSCP} to read the CPU counter). The interval between consecutive requests reflects the cache latency. When the interval is increased, the victim is supposed to have cache transactions concurrently. The attacker repeatedly visits the $EV$ for $x$ times (we set $x$ to 20 during the experiment) to obtain a sample for an interval, and analyzes the interval sequence to infer the access patterns of the victim applications (detailed in Section~\ref{sec:analysis}).

%The attacker records the timestamp when her core completes 10 rounds of traversing $U$. The interval between consecutive timestamps can be regarded as the delay for those L2 cache misses. 
%Besides, the increase of the delay can be attributed to the mesh bandwidth sharing with the victim. 
%Therefore, the victim's cache transactions incurs fluctuation in the delay sequence measured by the attacker.

%\subsection{Attack Optimizations}
%\label{subsec:optimization}

\zztitle{Using Huge Page} 
Huge pages are used to find EV easier. A recent work ~\cite{vila2019theory} proposes to construct eviction sets without huge page, which may help us relax the huge page requirement.

\begin{algorithm}[ht]
\SetAlgoLined
\KwResult{IntervalSeq }
\eIf{virtualized}{
    best\_path = rand\_path()\;
}{
     victim\_core\_ID = get\_core\_ID\_by\_PID(victim\_PID)\;
    best\_path = select\_best\_path(victim\_pos)\;
}

 L2\_set\_index = rand()\;
 \For{i in 3* (num of LLC slices)}{$EVs$.append( get\_EV( L2\_set\_index))\;}
 
 //Select an EV\;
 set\_affinity(best\_path.dst)\;

 \For{each $EV$ in $EVs$}{
    access(first addr of $EV$)\;
    \If{access time $\leq$ TH}{
    
        break\;}
}

//Start attack\;
set\_affinity(best\_path.src)\;
 \While{True}{
    \For{i in range(20)}{
        access($EV$)\;
    }
    IntervalSeq.append(access\_time)\;
 }
  \caption{The pseudo-code of \attack{} probe}\label{alg:flow}
\end{algorithm}

\subsection{The Pseudo-code}
\label{subsec:algorithm}

Here we summarize all steps covered in this section and show them in Algorithm~\ref{alg:flow}. To notice, the process of $EV$ generation is adapted from~\cite{yan2019attack}, by setting the size of the set of candidate $EV$ ($EV$s) three times as the number of LLC slices, in order to increase the chance of getting an $EV$ for a designated LLC slice. We verified the generated EV indeed belongs to the same cache set and LLC slice with the help of PMU.

%\zl{need explanations, why 3*num, } 
%Since \texttt{find\_EV} outputs an $EV$ for a random LLC slice, we generate $EV$s of three times of the number of LLC slices, hoping that later we have a high chance to get a $EV$ for our designated LLC slice.

%\zl{write a pseudo-code block for all steps in this section, including path selection, EV, optimization, ...}

\ignore{
\begin{itemize}

 \item Core to UPI. When all caches missed for a cache line that belongs to the CPU on another socket (\ie remotely homed), the home CPU will be responsible for fetching the cache line from the correct place (either a LLC slice, or L2 of a core, or memory), and then sending the data along the mesh, UPI stop, the mesh of the requesting CPU, until the cache line reaches the requesting core.

 \end{itemize}
} 
\section{Analysis of the \attack{} Side-channel}
\label{sec:analysis}

In this section, we provide quantitative analysis about the assumptions made in the previous sections. We first assess whether mesh congestion introduces measurable delays. Then, we analyze the impact of core locations on the attack program. Finally, we assess the capacity of the proposed side-channel.

%In this section, we analyze the side channel and evaluate the performance of the side channel, according to a series of measurement.

\subsection{Impact of Mesh Congestion}
\label{subsec:mesh_impact}

The key assumption that \attack{} can succeed is that the probe delay increases when the mesh is congested. We try to validate this assumption and uncover the root cause with an experiment.

%To make clear the root reason that causes the mesh delay increases and congestion, we design a experiment to monitor the internal of mesh to understand the reason.

\zztitle{Settings} We emulate a victim by using a process to access memory for 30us and then rest for 30us periodically. We place our victim at tile 0 and let it access memory addresses associated with tile 21, which generates mesh traffic along the top horizontal row. To monitor the internal of the intermediate mesh stops, we place 3 probes at the top horizontal path of the mesh structure, \ie, from tile 2 to 17, from 17 to 2, and from 7 to 12. Then, we turn on and off the victim program and record the PMU events respectively.

\begin{table}[]
    \centering
    \begin{tabular}{c|c|c }
         Events  &  w/o victim   & w/ victim  \\ \midrule[1.5pt]
         HORZ\_RING\_AD\_IN\_USE &  1004151  &  822149631   \\ \hline
         HORZ\_RING\_AK\_IN\_USE &  1952646432  &  2530753880  \\ \hline
         HORZ\_RING\_BL\_IN\_USE &  3914059524   &  4665246823   \\ \hline
         HORZ\_RING\_IV\_IN\_USE &  35639    &  102218   \\ \hline

         % TxR\_HORZ\_CYCLES\_NE&  &   &      \\ \hline
         TxR\_HORZ\_OCCUPANCY & 4414430426  & 5346595687  \\ \hline
         AG1\_BL\_CRD\_OCCUPANCY &  60996  &  482979 \\ \hline
         RxR\_OCCUPANCY &  4690   &  141706754 \\ \midrule[1.5pt]

         \makecell{STALL\_NO\_TxR\_HORZ\_ \\ CRD\_BL\_AG1} & 350 & 5005 \\ \hline
         RxR\_BUSY\_STARVED  & 9227  &   41261801    \\ \hline
         RxR\_CRD\_STARVED   & 2041  &   65177026    \\ \hline
         TxR\_HORZ\_STARVED  &  0 &  6942   \\ \midrule[1.5pt]

         TxR\_HORZ\_NACK    &  38951  & 188482   \\ \hline
         TxR\_VERT\_NACK    &  2      & 19405478 \\ \hline

        % TxR\_VERT\_ADS\_USED &  &  &\\ \hline

    \end{tabular}
    \caption{PMU events that are changed most rapidly.}
    \label{tab:pmu_comp}
\end{table}

\zztitle{PMU event comparison} Table~\ref{tab:pmu_comp} lists the PMU events that are changed the most among all the events we monitored. As we can see, the PMU events grow rapidly after we turn on the victim process. The increased events mainly fall into three categories: cycles in use, resource starvation and NACKs. The first category indicates that some components inside the mesh stop become busier, as the growth of these event counters means that a component costs more cycles in busy state. For example, events the increase of HORZ\_RING\_XX\_IN\_USE  indicates that more uncore cycles of the ring have been spent; events end with \_OCCUPANCY suggests more buffers or credits have been occupied.

The second category indicates that some components get stalled more frequently, because of lacking resources. For example, the increase of STALL\_ NO\_ TxR\_ HORZ\_ CRD\_ BL\_ AG1 means that more Egress buffer of BL ring of agent 1 was stalled during waiting for a credit. A mesh stop in the stall state would back pressure the sender, which increases the delay.

The third type indicates that more packets are lost due to congestion. For example, TxR\_HORZ\_NACK counts how many Egress packets have not been responded to the horizontal ring and NACKs were received. Packet loss would sharply increase the mesh delay. Mesh employs credit-based flow control, which assumes no packet would be dropped. Packet loss costs the stop considerable time to recover.

Based on the observation, we conjecture that when the two mesh flows (\ie, attack and victim) go through a mesh stop, components become busier, or even stalled, resulting in packet losses, which increases the delay of cache transactions.

\subsection{Impact of Tile Location}
\label{subsec:location}

In Section~\ref{subsec:path}, our theoretical analysis shows T1-T3 take the major share of mesh traffic, and we attempt to verify it here.
In the virtualized setting like cloud, the physical core ID is invisible to attackers. In this case, the attacker is unable to select the best path, even when the attacker knows the model and layout of the CPU. Below we show though information leakage is less acute, it is still significant under \attack.
%In this section, we show that the attack can still succeed when the probe is placed at a random tile.

\zztitle{Settings} We use Signal to Noise Ratio (SNR) of the square-wave alike interval sequence to quantify the information gain of the \attack{} channel. To learn SNR, we first convert the interval sequence from the time domain into frequency domain with a simple Fast Fourier Transform (FFT). We denote the magnitude value at the square wave frequency (\ie, 5kHz) as signal strength and the average magnitude at other frequencies as noise strength. SNR is their ratio. We fix the victim to tile 9 and assign the attacker to each other tile to obtain its SNR.

\zztitle{Results} We consider the max and median SNR over all paths for each attacker tile. Table~\ref{tab:loc_impact} shows our results. Tile 14 yields the best SNR, which aligns with the analysis of Section~\ref{subsec:path}. Specifically, T2 should be the major traffic source, as the right side of the victim tile has the most tiles, and tile 14 indeed carries most of the traffic. For other tiles, the SNR values are sufficient for recovering 1 bit, suggesting \attack{} is potentially effective even when the key path cannot be selected. In Section~\ref{subsec:sliding}, we evaluate this hypothesis.
%Place the figure there.

\begin{table}
  \centering
  \begin{tabular}{|c|c|c|c|c|c|}
     \hline
     UPI  & PCIE & PCIE  & RLINK & UPI2 & PCIE \\ \hline
     \makecell[c]{ \textbf{0} \\ 27.79  \\ 9.74}     &  \makecell[c]{ \textbf{2}\\21.03\\10.30}   & \makecell[c]{ \textbf{7}\\30.34\\11.74}   & \makecell[c]{ \textbf{12}\\23.42\\11.92}    & \makecell[c]{ \textbf{17}\\15.74\\10.24}   & \makecell[c]{ \textbf{21}\\18.61\\10.67}   \\
     \hline
     IMC0 & \makecell[c]{ \textbf{3}\\21.82\\11.63}    & \makecell[c]{ \textbf{8}\\19.28\\11.30}     & \makecell[c]{ \textbf{13}\\17.17\\9.60} & \makecell[c]{ \textbf{18}\\10.89\\7.31}   & IMC1 \\
     \hline
     \makecell[c]{ \textbf{1}\\15.60\\9.52}    & \makecell[c]{ \textbf{4}\\16.92\\11.77}    & \makecell[c]{\textbf{9}\\ Victim}     &  \cellcolor{yellow} \makecell[c]{ \textbf{14}\\49.75\\26.25}  \cellcolor{yellow}  & \cellcolor[gray]{0.8} & \makecell[c]{ \textbf{22}\\23.18 \\ 10.37} \\
     \hline
     \cellcolor[gray]{0.8}    &\makecell[c]{ \textbf{5}\\17.32\\12.17} & \makecell[c]{ \textbf{10}\\15.42\\7.32}  & \makecell[c]{ \textbf{15}\\23.28\\13.21} & \makecell[c]{ \textbf{19}\\19.89\\12.43} & \makecell[c]{ \textbf{23}\\16.40\\8.39} \\
     \hline
     \cellcolor[gray]{0.8}    &\makecell[c]{ \textbf{6}\\16.47\\11.46} & \makecell[c]{ \textbf{11}\\24.25\\11.77}  & \makecell[c]{ \textbf{16}\\21.57\\9.65} & \makecell[c]{ \textbf{20}\\19.35\\10.58} & \cellcolor[gray]{0.8}  \\
     \hline
   \end{tabular}
  \caption{Layout of Xeon Scalable 8260 CPU with the measured SNR. Numbers in every core die represents the Tile ID, maximum and median SNR values respectively. }\label{tab:loc_impact}
\end{table}

\subsection{Covert Channel Throughput}
\label{subsec:capacity}

%\zl{RA seems dislike this part very much, downplay it?}
Transient execution attacks~\cite{kocher2019spectre,lipp2018meltdown,van2018foreshadow} turned cache side-channel to covert channel for information exfiltration. Similarly, our stateless side channel could be re-purposed as a covert channel, and we are interested in its throughput.
%we characterize the resolution of the probe by constructing a covert channel and measuring the channel throughput.
Hence, we construct a sender and the pseudo-code is shown in Appendix~\ref{appendix:covert}.
%\zl{show the pesudo-code of your sender program, or in Appendix}.
In essence, it repeats the process of accessing an $EV$ for $\frac{T}{2}$ (turn on the mesh traffic) and then running NOP loops for $\frac{T}{2}$ (turn off the mesh traffic), assuming $T$ is the execution time. The receiver program runs the \attack{} probe to collect a sequence of intervals and decodes the patterns.

%We expect the intervals to be prominently increased when congestion happens and decreased when congestion ends

%The capacity of the side channel can be estimated by the minimum pattern that can be recognized by the probe. To measure the amount of information can be leaked with our attack,
%we design a simple toy victim that produces on-off style memory access pattern and try to attack the toy victim to see how fast the memory access timing can be learnt.

\begin{figure}[ht]
    \centering
    \includegraphics{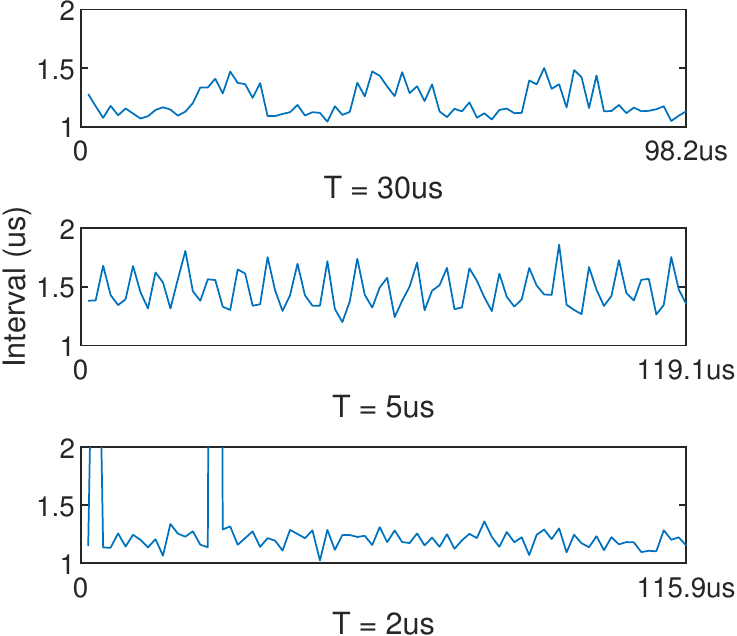}
    \caption{Interval sequence collected by the receiver of different $T$. The x-axis is the running period of the sender. The three sequences have the same number of sample points.}
    \label{fig:square}
\end{figure}

We run the experiment with different $T$, and Figure~\ref{fig:square} shows the interval sequence observed by the receiver, for a period of around 100us. When $T$ is 30us and 5 us, we expect around 3 and 20 peaks to be observed, and the interval sequence proves it. However, when $T$ drops to 2us, we are unable to identify 50 peaks. As such, we conclude the channel capacity is around \textbf{200 kbps} (one bit per 5 us).
It also means that if a victim program accesses memory (including cache) at the latency of less than 5us, the attacker program cannot decode its access pattern.
%The capacity also implies that one can use the toy program to set up a 200 kbit/s covert channel between different cores in the same CPU.

\section{Attacking RSA}
\label{sec:evaluation}
%In this section, we show how an attacker can employ volatile cache attack to steal the RSA decryption key of a victim program.

In this section, we evaluate the step 3 (Secret Inference) described in Section~\ref{subsec:steps}. We choose two Java RSA implementations as the target, one is known vulnerable to the conventional timing side-channel attacks, and another is hardened with sliding window. We first describe the platform for the attack evaluation. Then, we elaborate the attack results and our findings.

\subsection{Experiment Settings}
\label{subsec:platform}

\begin{table}[h]
  \centering
  \begin{tabular}{|c|c|}
    \hline
    % after \\; \hline or \cline{col1-col2} \cline{col3-col4} ...
    CPU & Intel Xeon Scalable 8260 \\ \hline
    Main board & Super Micro X11SPA-T/TF  \\ \hline
    Memory & 64GB \\ \hline\hline
    OS & Ubuntu 18.04 \\ \hline
    JDK version & 11 \\ \hline
    CAT Version & 1.2.0-1 \\ \hline
  \end{tabular}
  \caption{Hardware and software for the attack evaluation.}\label{tab:exp_spec}
\end{table}

Table~\ref{tab:exp_spec} shows the hardware and software specification of our experiment platform. The CPU belongs to the latest generation Intel CPU, \ie, Cascade Lake-SP, which can be purchased from retailers. The layout of the CPU has already been reverse-engineered by us, showing in Table~\ref{tab:cha_cpu_pos_2}.
%\zl{Intel CAT}

To validate our claim that \attack{} bypasses cache partition, \textbf{we turn on Intel CAT for the entire experiment duration}. This is the same setting as the recent advanced cache attacks like Xlate~\cite{van2018malicious}. We did not enforce the stronger protection, temporal isolation~\cite{ge2019time}, because it builds on seL4. However, as \attack{} is based on the stateless channel, which admittedly is out of the mitigation scope~\cite{ge2019time}, we expect \attack{} is effective.
%\zztitle{Cache Isolation} A key innovation of volatile cache attack is that it defeats cache isolation protection. We assume the victim is under the protection of cache isolation.
For Intel CAT, we create two COS. The core running the victim program is bond with COS 1, while other cores are bond with COS 2, by using the command \texttt{pqos} of intel-cmt-cat package~\cite{pqos} (with parameter \texttt{-e}). COS 1 is exclusively allocated with one way of LLC while the rest 10 ways are allocated to COS 2 exclusively.
Therefore, the attacker program will not share with any L1/L2/LLC cache with the victim program.

We evaluate \attack{} against Java programs because
%such high-level programming languages have more visible memory access pattern related to sensitive data. For example,
Java yields more distinguishable patterns of mesh traffic, comparing to other languages without automated memory management, like C++.
Java Virtual Machine (JVM) creates a new object for the same variable for each iteration within the loop, and uses Garbage Collection (GC) to manage the old object, which makes RSA particularly vulnerable. Here we use the example shown in Appendix~\ref{appendix:rsa}.
For the line of $m = m^2 \; mod \; n$, the variable $m$ references two different objects before and after its execution.
The allocation of the second object causes cache misses, which introduces more memory read/write over mesh interconnect.
In contrast, in languages like C, the developer may reuse the object to save memory, in which case the memory access may be completed within the private cache of the core.

%In this case, there would not be any process, including the attacker's, that shares LLC with the victim.
%The attacker will bind her program in a core but never the one in COS 1. Therefore, the attacker does not share any private cache with the victim. In total, there is no any cache shared between the victim and the attacker.

%with fast modular exponentiation and sliding window

We assume a 2048-bit private key is used. The private key consists of an exponent $d$ and a modulus $n$. The message to be decrypted is segmented into 2048-bit groups ($m$). Therefore, $d$, $m$ and $n$ are all 2048 bits. The attacker aims to infer the bit sequence of $d$.

\subsection{Attacking Fast Modular Exponentiation}
\label{subsec:toy}

%We use RSA as a victim program to evaluate the effectiveness of \attack. Similar to previous cache side-channel attacks~\cite{yarom2014flush+,liu2015last,yan2019attack}, we assume the code has a vulnerable implementation that yields different patterns for key bit 0 and 1. The code implements the fast modular exponentiation algorithm, as shown in Algorithm~\ref{alg:rsa}, for RSA decryption.

We implemented the basic fast modular exponentiation (Algorithm~\ref{alg:rsa}) in Java following the code found in~\cite{bouffard2014generic}. As described in Section~\ref{subsec:rationale}, different key bit leads to different memory access patterns.
This algorithm is also implemented in GnuPG 1.4.13, which is evaluated by other cache side-channel attacks (\eg, ~\cite{yan2019attack}).
We tested 100 different keys with this RSA program as victim, and for each key, we let the program run 20 times. Hence, there are in total 2,000 traces collected by the \attack{} probe.

\zztitle{Selection of Key Path}
We simulate the non-virtualized setting here, of which the attacker learns the core ID. We assume the victim program runs in a core randomly assigned by the OS and the attacker can select a core among the rest to construct key paths. In Section~\ref{subsec:sliding}, we simulate the virtualized setting, and obscure the core ID.

According to Section~\ref{subsec:path}, the attacker can estimate the traffic reaching/leaving the victim core. Though the attacker has the freedom to select multiple paths based on the estimation, we found one vertical path is sufficient to collect clear-enough signals about the victim RSA program. Therefore, we select the top-ranked vertical path and place the attacker program to one end of the path.

%\zl{I feel this part isn't good. It means the key path selection is useless.}

%After having analyzed the the RSA program, we found both the horizontal and vertical paths carry large mesh traffic.
%Placing probes on both paths yields similar results. Besides, as mentioned before, it's enough for the attacker to place only one thread to collect clear enough sequence.
%Therefore, we assume the victim is running at random core and the attacker place only one probe on the longest vertical path to collected data.

\begin{figure}[ht]
    \centering
    \includegraphics[width=0.4\textwidth]{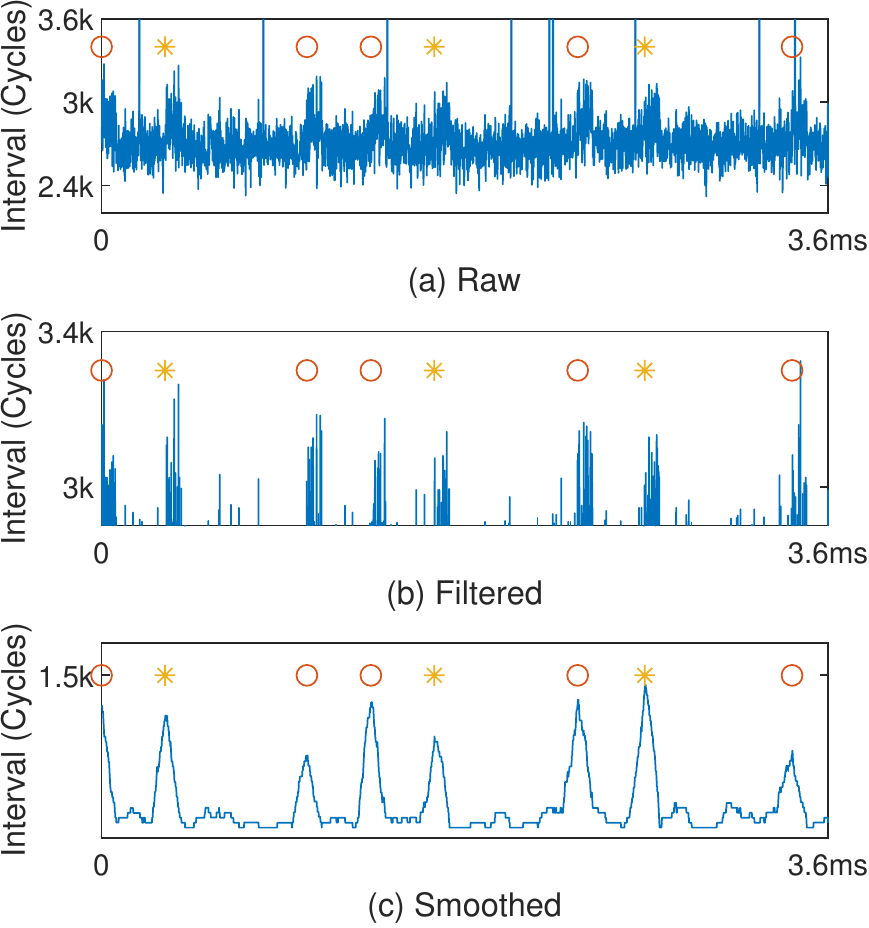}
    \caption{(a) The raw interval sequence collected by attacker's probe. (b) The sequence after filtering. (c) The sequence after smoothing, with circles and stars representing the starting time of bit 0 and 1. The key stream is \texttt{01001010}. The max value of y-axis of (c) reduces to 1.5k because smoothing averages the values in a window. }
    \label{fig:rsa}
\end{figure}

\zztitle{Analyzing the Interval Sequence}
\label{subsec:sequence}
Figure~\ref{fig:rsa}a shows the interval sequence mapped to the first 8 bits of a $d$, which is \texttt{01001010}.
Differentiating bit 0 and 1 turns out to be more challenging than the stateful cache attacks, because their probe tracks the cache access of each function (\eg, \texttt{sqr} and \texttt{mul}~\cite{yan2019attack}), while \attack{} works at the granularity of a program. Yet, with the heuristics described below, bit 0 and 1 can be differentiated.

Specifically, when 0 is encountered, a sharp rise will be observed, which we call \texttt{A} rise. When 1 is encountered, \texttt{A} rise will be observed first, followed by one or two smoother rises, which we call \texttt{B} rise. Besides, it takes longer time to process a bit 1 than bit 0.

As \texttt{A} rise is more discernible than \texttt{B} rise and other irrelevant data points, we use a threshold $(2900, 4200)$ to keep \texttt{A} rises first. Figure~\ref{fig:rsa}b shows the data points after filtering. After then, the data points are smoothed, \ie, taking the average of the points in a window, so \texttt{B} rises are expected to be diminished and \texttt{A} rises become more distinguishable, as shown in Figure~\ref{fig:rsa}c.
With the smoothed sequence, the attacker starts to find the peaks over 600 cycles, which are expected to be \texttt{A} rises. Then, she examines the peaks in the execution time window of bit 0 and 1 close to a \texttt{A} rise respectively. %In particular, she sequentially finds peaks in the time windows of bit 0 and 1 respectively.
The window with a higher peak indicates which bit it is.

%\zl{i still don't understand following sentences}
%If only one peak found at a window, we label the bit accordingly. Otherwise, if both windows have peaks, we label the bit according to which peak is higher. If no peak found at either window, we abandon the peak and restart at the next point.

%She finds a climax with enough height as a A rise and then try to see if there is a climax at the next potential bit 0 position and bit 1 position. If only one position has climax, a decision can be made. If no climax found on both positions, the attacker can abandon this bit and continue to find the next A rise, hoping that later error correction can correct the bit. %If climax found on both positions, she try to make decision according to which climax is higher.

%The attacker can easily distinguish bit 0 and 1 with the interval between consecutive A rises. A challenge is that some B rises may be mistaken as A, as some A and B are pretty similar.

%To overcome the challenge,
%the attacker can firstly  set a range that some A rises fallen into, but B rises as well as other abnormal rises can hardly. With this range, the attacker can filter out those sample points out of the range and only consider the sample points inside the range (see ).

%As we can see, the range still cannot filter out all B rises, but is much more clear. The attacker can then smooth the sequence such that B rises are diminished (see Figure~\ref{fig:rsa}.c), after which the B rise is much shorter than rise A. Then the attacker can start to make decision on the smoothed sequence.

\begin{table}[ht]
    \centering
    \scalebox{0.9}{
    \begin{tabular}{|c|c|c|c|c|c|c|c|} \hline
    & \multicolumn{5}{c|}{Edit Distance} &  \multicolumn{2}{c|}{Avg. LCS} \\ \cline{2-8}
           & $\leq$10 & $\leq$50 & $\leq$100 & $>$100 & Avg. & Str & Seq\\ \hline
       Smoothed  &17&67& 2 &14 & 15.4    & 760.6       & 2039.5 \\
Corrected   &47&9& 2  & 25& 7.8   & 2040.0      & 2040.4 \\ \hline
    \end{tabular}}
    \caption{Results of the RSA key recovery experiment (Fast Modular Exponentiation). ``LCS-Str'' means the longest common sub-string. ``LCS-Seq'' means the longest common sub-sequence.  Average metrics for the smoothed cases are computed on the inferred keys whose edit distances are less than 50. For the error-corrected case, we choose 10.}
    \label{tab:rsa}
\end{table}

\zztitle{Results}
\label{subsec:effectiveness}
%The attacker has very high chances to learn the whole private key, according to our results.
%The cell of edit distance reports the number of cases whose edit distances fallen into the range. \eg, 67 under the cell ``$\leq 50$'' means 67 keys have edit distance below 50 but above 10.
%\zztitle{Evaluation Metrics and the Result on the Smoothed Waveform}
For each key used by the victim program, we compute the edit distances between all the inferred keys with the ground truth, and take the smallest value.
%If the value is less than 50, we regard the key recovery as successful.
Table~\ref{tab:rsa} row ``Smoothed'' shows the distribution of the edit distances. Out of the 100 keys, for 17 keys the edit distances are at most 10, 84 (17+67) are at most 50. We look into the 84 keys, and compute the average edit distance and two other metrics:
the longest common sub-string (LCSStr) and the longest common sub-sequence (LCSSeq).
%The values are 15.4, 760.6 and 2039.5 respectively.
%\zl{I think smallest is unfair.}
%Besides, those failed cases still leak large chunks of correct bits.
The result suggests a large portion of the key has been inferred. For instance, the average LCSStr is 760.6, meaning that a chunk of 760 consecutive bits can be precisely recovered.

%\zztitle{Coarse Recovery}

%\zztitle{Error Correction}
Moreover, we found the inference accuracy can be enhanced with error-correction techniques. We chose De Bruijn graph~\cite{lin2016assembly}, a technique widely used to correct gene sequence errors, for this task. In essence, for a group of long sequences, it breaks each one to sub-sequences and drops the less frequent ones. Then it concatenates the remaining sub-sequences back to a complete sequence.
%\zl{can you summarize what this graph tries to do}

With the De Bruijn graph, for a group of 20 inferred keys, we can correct the errors and generate 1 key. We compare each generated key to the ground truth, and the row ``Corrected'' of Table~\ref{tab:rsa} shows the result. This time, 47 inferred keys have an edit distance less than 10, and we further compute the average edit distance, LCSStr, and LCSSeq for them. It turns out the average LCSStr can be as high as 2040, meaning that \textbf{only 8 bits are incorrectly predicted}.
%Besides, it is very close to LCSSeq, meaning that the 2040 bits are consecutively and correctly predicted.
Since LCSStr is highly close to LCSSeq in this case, the errors are expected to be distributed in the head and tail of the key stream. To recover the exact key, the attacker only needs to enumerate $9*2^{8}$ combinations of bits.

%\zl{I don't understand}
%\zl{i'm quite confused about you metric, why not count the corrected version using 50 E.D.?}

%\zztitle{Performance Impact on the Victim Program}
%\attack{} contends the mesh link with a victim program, which should enlarge the execution time of the victim program. However, we found the extra overhead is quite small.
%The performance degradation made by the probe to the victim is negligible.
%When \attack{} is running, the execution time for a round of RSA decryption increases from 1022 ms to 1049 ms only.

\subsection{Attacking Sliding Window RSA}
\label{subsec:sliding}

Fast Modular Exponentiation was known to be vulnerable to timing side-channel attacks. To defeat timing attacks, the recent RSA implementations have been upgraded to use the \textit{Sliding Window} algorithm, which decouples key stream from \texttt{mul}/\texttt{sqr} execution sequence. For example, the Crypto suite of JDK (\ie, \texttt{javax.crypto.} \texttt{Cipher}) uses Sliding Window, and GnuPG has adopted Sliding Window after version 1.4.13.

However, a recent work~\cite{bernstein2017sliding} showed that \texttt{mul}/\texttt{sqr} execution sequence can still be utilized to crack RSA that is based on Sliding Window. Given a  \texttt{mul}/\texttt{sqr} sequence, their algorithm (\textit{Sliding right into disaster}, or \textit{SRID} for short) is able to either output the 100\% correct inference for a key bit (\ie, either 0 or 1), or output X, meaning the algorithm is unable to get a correct inference.
With SRID to crack 2048-bit RSA key, 5-bit sliding window RSA implementation leaks 33\% of the key bits.
JDK uses 7-bit sliding window, in which case around 30\% bits is expected to be recovered theoretically.

\attack{} can make full use of SRID to recover key bits. The attacker needs only firstly recover \texttt{mul}/\texttt{sqr} sequence with the similar approach as Section~\ref{subsec:sequence}, then SRID is applied to output 0, 1 and x.

%we find there are three kind of patterns from which we can tell the multiply and square operation, such like the probe are congested when we meets multiply operation or square operation. If we get the sequence of multiply, we can deduce the concrete time of square,vice versa. To illustrate the effectiveness of our probe, we write scripts to extract the whole \texttt{mul} sequences from one pattern and deduce \texttt{sqr} time by time gap.

%The accuracy of extraction is easy to achieve 100\%. Then she applies the rules in~\cite{bernstein2017sliding} to infer part of the key streams.

\zztitle{Settings} We tested the official JDK implementation of RSA (\texttt{javax.} \texttt{crypto.Cipher}) as the victim, and placed the program in a fixed tile. At the same time, we place probes at random positions to collect interval sequences and infer keys, \textbf{simulating virtualized setting where the attacker cannot use core ID to select the key path}. The experiments were repeated 1000 times.

We aim to recover \texttt{mul}/\texttt{sqr} from the interval sequence, which  is different from the analysis described in Section~\ref{subsec:sequence} because three new patterns have been observed, and none of them overlap with \texttt{A} and \texttt{B} rises. In Appendix~\ref{appendix:sw_seq}, we elaborate how to map them to \texttt{mul}/\texttt{sqr} sequences.
%\zl{can you fill the appendix? Done}

\zztitle{Results} Among all these 1000 traces, 32 exhibit obvious patterns, and \texttt{mul}/\texttt{sqr} were recovered \textbf{without error}.
Assuming the victim repeatedly runs RSA and the attacker keeps profiling mesh traffic, it will take on average 31 (1000/32) rounds to get a perfect \texttt{mul}/\texttt{sqr} sequence.
Then, we implemented the SRID algorithm and tested on the 32 \texttt{mul}/\texttt{sqr} sequences. In average, \textbf{31.05\% key bits} can be inferred. A recent work~\cite{Oonishi2020} also pointed out that the SRID algorithm has room to be improved for a better recovery ratio. Therefore, we believe more key bits can be recovered pending on the new algorithms.

%, indicating that the attack successful rate is 3.2\%. We recovery 3 random keys and compare it to the ground truth,and we recovery 611,563 and 599 keys respectively. The average recovery rate is 28.9\%.

%According to our experiment results, the \texttt{mul}/\texttt{sqr} sequences can be 100\% recovered. With the help of~\textit{Sliding right into disaster}, 28.9\% key bits are inferred in average.

\ignore{
\subsection{Attacking GnuPG}
\label{subsec:gnupg}

\begin{figure}[ht]
    \centering
    \includegraphics[width=0.42\textwidth]{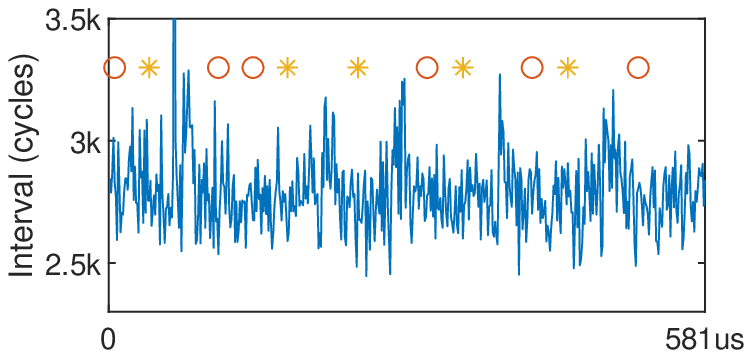}
    \caption{Interval sequence collected when attacking GnuPG 1.4.13. Orange circle is the start time for bit 0, while yellow star is the start time for bit 1. The stream is \texttt{0100110101}.}
    \label{fig:gnupg}
\end{figure}

Like previous works~\cite{yarom2014flush+, liu2015last, yan2019attack}, We also tried to attack GnuPG 1.4.13~\cite{gnupg1413}, which implements RSA in fast modular exponentiation but is vulnerable to timing analysis.
We use the 8192-bit RSA key\footnote{Supporting the 8192-bit key requires changing the key length limit constant of GnuPG 1.4.13.} since it enlarges the execution time of GnuPG and allows us to collect enough sampling points (\attack{} cannot capture memory/cache activities under 5us).

Figure~\ref{fig:gnupg} shows the interval sequence collected by the attacker when the stream is \texttt{0100110101}.
As we can see, the interval patterns of bits 1 and 0 are distinguishable. There is a high rise of intervals when bit 1 is encountered but the rise is not obvious for bit 0.
} 
\section{Discussion}
\label{sec:discussion}

\zztitle{Limitations}
%\zl{applications which do not generate a lot T1-T3 traffic, moved from Section~\ref{subsec:path}}
%Like previous works in cache side-channel attacks~\cite{yarom2014flush+,liu2015last,yan2019attack}, we attack the square-and-multiply exponentiation algorithm of RSA under \attack. There might be other applications whose data access is concentrated in the private cache, thus introducing much less mesh traffic. We acknowledge \attack{} might be ineffective in this case. We plan to profile more applications to understand which are vulnerable.
%According to our analysis, it extensively accesses LLC.
%complete execution on  only and do not emit much mesh traffic. In this case, the attacker may capture no cache activities. We do not consider methods to intentionally evict victim's cache lines to force the victim to produce mesh traffic, as we assume the victim is protected under cache isolation.
1) \attack{} attempts to congest T1-T3 traffic, but T4-T7 may also leak sensitive information about the victim application. For example, I/O intensive applications could introduce prominent mesh traffic between PCIe stop and LLC slices.
%The pattern of I/O traffic may also be related to sensitive information, which is also valuable to consider. Other traffics may also indicate some pattern of the victim, but without documentation, it's hard to profile. Therefore, we consider the traffic that between LLC, core and IMC, which predominates the mesh traffic.
We plan to test such applications in the future. 
2) \attack{} bypasses the existing defenses at the cost of obtaining coarser-grained information about cache activities. 
In stateful cache attacks like \textsc{Prime+Probe}, the attacker can precisely evict a cache set shared with a victim and then learn the \textit{which} memory address/range accessed by the victim.
However, \attack{} only tells \textit{when} the victim accesses memory/cache. 
As such, we choose the Java-based RSA implementations which leak more information.
%This also impacts our attack on GnuPG (when the key size is smaller, like 2048, the interval sequence is hard to decode).
%Yet, we found the leaked information is sufficient to conduct high-profile attacks.
3) Our evaluation is done on a single Xeon Scalable CPU. In the future, we plan to test other Intel CPUs and the ARM server CPU Neoverse~\cite{arm_mesh}.

%\attack{} exploits a vulnerability in the implementation of the fast modular exponentiation algorithm, which introduces different timing patterns for key bit 1 and 0. Previous cache side-channel attacks~\cite{yarom2014flush+,liu2015last,yan2019attack} exploited the same vulnerability in GnuPG 1.4.13, which is deprecated and has been patched. For the later patched implementations, \attack{} can be thwarted, together with other attacks.
%Still, we believe \attack{} has practical value, because it enlightens a new leakage channel that is overlooked previously. 

%Plenty of apps are not protecting their memory access pattern and could be the target of \attack{}.
%We didn't successfully attack commercial applications with RSA encryption. We tried to attack GnuPG but found there are already countermeasures to balance the time consumption for bit 0 and 1, defeating all timing attacks against RSA. 

%leaks the overall memory access pattern. The attacker does not know which line was accessed but can only learn whether there is memory accessed. 

%Mesh is not only adopted by Intel. Amazon server CPU Neoverse also comes with mesh. We would design experiments to check if \attack{} can be launched on Neoverse.

\zztitle{Implications of \attack}
%\zl{talk about why it is hard to defend? why can't we turn off sharing of mesh routes/CPU?}
The key take-away message from our study is that the cache side-channel attack can be done without changing the microarchitectural state through the stateless interconnect. This is counter-intuitive at first sight, but the new interconnect design intertwines the cache lines \textit{on the move} from different applications, introducing new types of resource contention that enables our attack. 

%Though the evaluation result of \attack{} indicates the interconnect should be protected. 
To prevent the interconnect from leaking an application's status, the cache traffic could be regulated under the \textit{non-interference} property~\cite{goguen1982security}. Previous works verified the code of programs to detect the vulnerable ones that violate this property~\cite{chen2017precise}, but doing so on the interconnect traffic has to model the highly complex microarchitecture. 
Like cache partition, if interconnect bandwidth can be partitioned, \attack{} might be thwarted. However, as mentioned in~\cite{ge2019time}, ``no support for bandwidth partition exists on contemporary mainstream hardware''. Though Intel recently proposed a technique named Memory Bandwidth Allocation (MBA)~\cite{intel_mba}, which limits the bandwidth a core can issue to memory. The limit is an approximation and insufficient for threat mitigation~\cite{ge2019time}.

%Given most of the microarchitectural implementations are proprietary, mitigation is non-trivial.
%\attack{} is hard to defend, because it only requires the mesh interconnect to be shared while does not assume any cache share between victim and attacker. Isolating cache for different cores does not prevent the traffic co-location on mesh.

%\zl{mention that there is no hardware support for bandwidth partition, and Intel MAB is not good enough~\cite{ge2019time}}

%\textit{non-interference} property actually go away from the design of mesh interconnect: disaggregating large LLC into independent slices that can be accessed concurrently.

\zztitle{Potential Defenses}
%The mesh interconnect can not be easily split and isolated for different cores. Cache can be easily split by ways and assigned to different cores. If a mesh link is disabled for a core, the core would be disconnected from LLC slices, memory controllers (IMC stop) and PCIE that uses the link. Without the ability to access one single LLC slice or IMC, the core would be unable to access the complete memory space.
Instead of strong mitigation based on the non-interference property or partition, we believe mechanisms that increase the attack difficulty is more likely to be adopted. We discuss a few directions that might lead to a successful defense.

%Though \attack{} is hard to defend, we propose some possible countermeasures for CPU vendors, expecting it would be more difficult for attackers to launch \attack{}.

One important prerequisite of \attack{} is that when a mesh stop is congested, the packets through it would not be detoured. Under the simple YX routing, the routes between two tiles are fixed, easily satisfying the attack prerequisite. Horro \etal~\cite{horro2019effect} explored dynamic routing to optimize the coherence traffic. Though their goal is to reduce the access latency, it has the potential to break the attack prerequisite.
%Dynamic routing could mitigate the attack. Current mesh implementation uses the , in which the route is fixed, leaving opportunities to attackers to know where the victim traffic would be. If a core can select a path to get around the attack traffic, the probe may no longer be able to co-locate with the victim traffic. Dynamic routing would make the design of mesh more complex but it is worthwhile, as dynamic routing may increase the throughput of mesh interconnect with higher utilization of mesh.

\attack{} mainly targets the cross-tile T1-T3 traffic related to LLC. Intel CAT allocates different cache ways to different applications to implement cache isolation, but it does not partition the cross-tile LLC accesses. If cache partition can be done at the level of LLC slice, \eg, placing the data frequently accessed by an application to the LLC slice local to its occupied cores, \attack{} might be deterred. Farshin \etal~\cite{farshin2019make} designed a slice-aware memory management mechanism and showed it can realize cache partition, which holds promises. 
%Intel may also consider to allow managers to isolate cache by slices rather than by ways. 
%If the attacker's cache access can be restricted to her own cache slices in which no mesh path should be  shared with victim, the attacker would be unable to co-locate with the victim on mesh. 
On the other hand, such a mechanism does not prevent attackers who exploit T4 -T7, \eg, the traffic associated with IMC. Hence, a new memory management mechanism might still be needed.

%The sub-NUMA cluster design of Intel CPUs can split one CPU into two NUMA nodes, each of which exclusively uses an IMC, but it does not prevent a core from accessing LLC slices at the remote NUMA node. 
%Intel may modify the design to also restrict the access to LLC slices.
Cache randomization is another potential direction. By forcing the mapping between physical addresses and cache set index dynamic and unpredictable, finding the right eviction sets is expected to be more difficult, which could make the probe of \attack{} unstable. However, recent studies~\cite{song2020randomized,bourgeat2020casa,purnal2021systematic} have shown the state-of-the-art approaches like CEASAR-S~\cite{qureshi2019new} and ScatterCache~\cite{werner2019scattercache} are broken under new attack methods. Though Song \etal\ proposed a fix to address the new attacks~\cite{song2020randomized}, the defense is demonstrated on a RISC-V simulator. When the commercial processors adopt such defenses is yet unknown.

%Another direction could be cache set randomization. Current CPUs directly use part of physical addresses as cache set index, which means the mapping between addresses and cache set is static. Recent research tried to ~\cite{song2020randomized}, which stops the attacker from generating stable probes.
%\zl{we should talk about the oakland'21 papers about cache randomization.}
%\input{tex/Related.tex}
\section{Conclusion}
In this work, we reveal stateless cache side-channel, or \attack{}, that can leak memory access patterns of a victim program, by exploiting the traffic contention on the CPU mesh interconnect. The side-channel is different from previous cache side-channels, in that it does not rely on stateful micro-architectural state changes made by the victim. Therefore it can bypass both spatial and temporal isolation. 
To reveal the consequences of \attack{}, we conducted an analysis on RSA implementations to infer the private key from the collected delay sequences. 
The results show that \attack{} is very effective. We believe mesh interconnect opens up new opportunities for security research, and its implications should be further examined.

%First step, more to uncover.

\bibliographystyle{ACM-Reference-Format}
\bibliography{main}

\appendix

\section{Intel Skylake-SP cache Spec}
We introduce the cache specification of Intel Xeon Scalable family processors, which we used as our platform.

\zztitle{Address Mapping} The lowest 6 bits reflect the block offset within a cache line. The bits in the middle indicate the index of the cache set containing the line (bits 15:6 for L2 and 16:6 for LLC). The upper bits form a cache \textit{tag}, which indicate whether the data is in the cache.

As LLC requests are all managed by CHA inside a core, for a LLC access, CPU has to decide which CHA to talk to. The decision is based on a proprietary hash function which is not fully reverse-engineered yet~\cite{yan2019attack}.

\zztitle{Cache Structure} For Skylake-SP processors, LLC is designed as \textit{non-inclusive} to the private caches. Before Skylake-SP processors, LLC is \textit{inclusive}, meaning that a cache line in L2 cache has a replicate in LLC. For non-inclusive LLC, a L2 cache line may or may not have replicate in LLC, which is determined by the cache eviction policy. As a result, Xeon scalable CPUs have much larger effective cache size (the sum of L2 and LLC) compared to the previous generations (LLC only).

%\zl{appendix?}

%All cache lines in the same sub-space has the same hash value over part bits of their physical addresses.
%Inside the sub-space, addresses are mapped into sets according to the middle 11 bits (see Figure~\ref{fig:address}) \footnote{summarized the hash function for determining CHA \# has not been fully reverse-engineered by \cite{yan2019attack} and .}.
%Each CHA is accompanied with a LLC slice and handles cache transactions received from mesh.

%, because a CPU can hold cache of size up to , while previous generation only hold up to the size of L3, as L2 lines are only replica of L3 lines.

%Along with and thanks to mesh, Intel redesigned the cache system of Xeon scalable processors, which comes with larger overall size and better scalability. Mesh provides more powerful switching capabilities between cores, comparing with previous generations. With the help of better switching capability, Intel was able to split the last level cache (LLC) into more slices and distribute them in a wider area to have larger size. Specifically, cache lines in L2 is no longer required to be present in L3.

%A key change from Skylake-SP on is the use of non-inclusive LLC. Previous Intel Xeon CPUs' (\ie before skylake-SP) cache system consists of three level caches that are all inclusive. When it comes to Skylake-SP, LLC is non-inclusive, while L2 remain inclusive.

\begin{figure}[ht]
\centering
    \begin{tabular}{|c|c|c|c|} \hline
    \quad\quad\quad 63:17 \quad\quad\quad & 16 & 15:6 & 5:0 \\ \hline
    \multicolumn{2}{|c|}{L2 Tag} & L2 set index & \multirow{3}{*}{Offset} \\ \cline{1-3}
    LLC Tag &  \multicolumn{2}{|c|}{ LLC set index  } & \\ \cline{1-3}
    \multicolumn{3}{|c|}{Hash to LLC slice ID}& \\ \hline
    \end{tabular}
    \caption{Mapping between memory address (physical) and cache.
    %\zl{appendix?}
    %\zl{change L3 to LLC}
    }
    \label{fig:address}
\end{figure}

\begin{table}[ht]
\centering
\begin{tabular}{c|c|c|c}
  \hline
    & Size & Associative & Set\\ \hline
  L1-I & 32KB & 8-way & 64\\
  L1-D & 32KB & 8-way & 64\\
  L2  & 1 MB & 16-way & 1024\\
  LLC slice & 1.375MB & 11-way & 2048\\
  \hline
\end{tabular}
\caption{Cache configuration for Skylake-SP, Cascade Lake-SP and Cooper Lake-SP CPU families~\cite{xcc_wiki}. L1-I and L1-D are for instruction and data separately.
%\zl{appendix?}
}
\label{tab:cache_config}
\end{table}

%\zztitle{Cache Access}

\begin{figure}[h]
    \centering
    \includegraphics[width=0.45\textwidth]{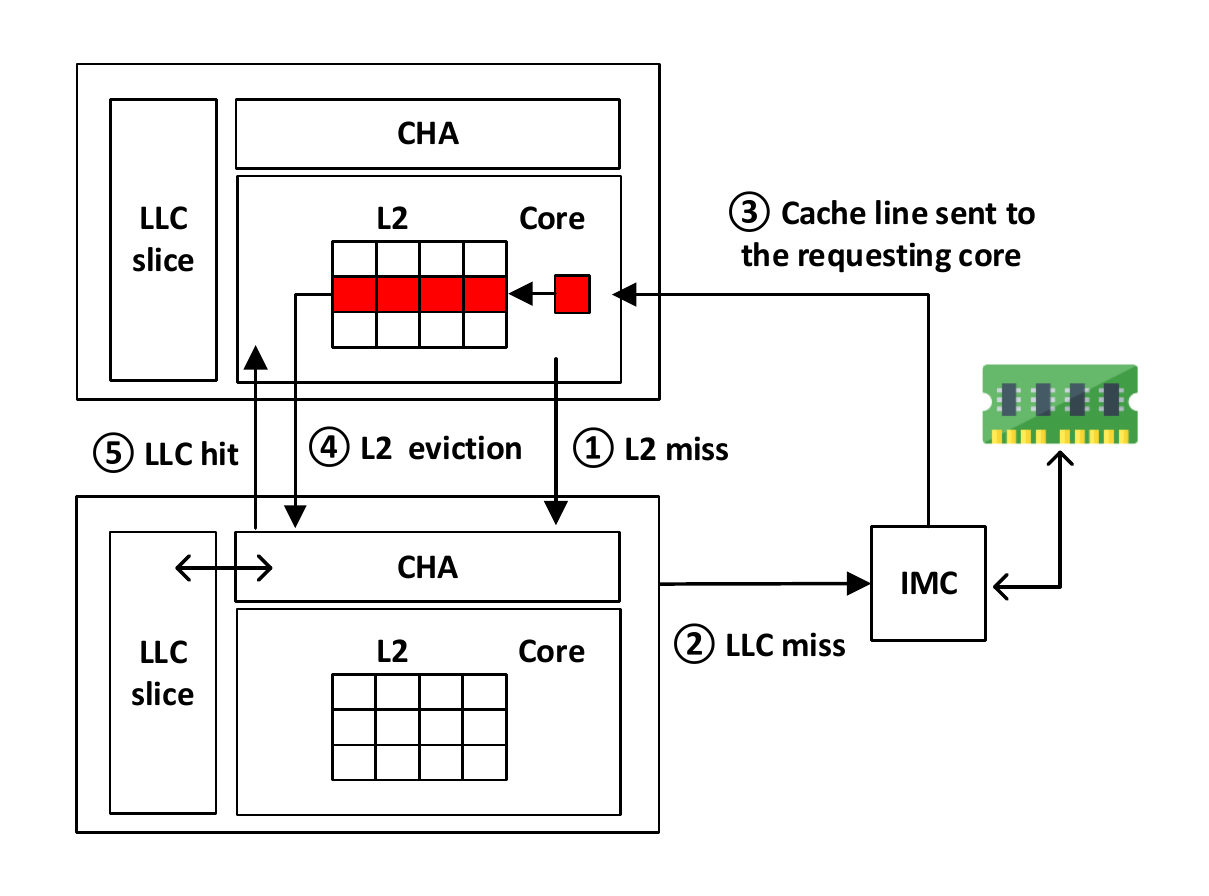}
    \caption{The process of the cache access.
    %\zl{can you redraw it?}
    %\zl{can you add some text to each step? it's better to draw two cores, CHA, LLC, L2, memory and mesh}
    }
    \label{fig:cache_trans}
\end{figure}

\ignore{As the same memory block can be placed in multiple places in the cache hierarchy and the same cache line can be read/write by different cores, the cache should be kept coherent to avoid the access of the outdated data.
To this end, Intel proposed the \textit{MESIF (Modified, Exclusive, Shared, Invalid, Forward)} protocol~\cite{} for cache coherence, and we describe the five cache states managed by MESIF below.
``Modified'' means the line is in the private cache of the owing core and is dirty as it has been written.
``Exclusive'' indicates the line is stored in a single core.
``Shared'' means the line is potentially shared by multiple cores.
``Invalid'' means the line is not in cache.
``Forward'' suggests the core holding the line is responsible to forward the line to cores asking to share the line.

Though how MESIF is implemented in an Intel CPU is not officially documented, through the reverse-engineering efforts~\cite{yan2019attack}, the details are partially revealed. }

\zztitle{Cache Access}
\label{sec:access}
Here we describe the process of cache access, focusing on the non-inclusive LLC (also illustrated in  Figure~\ref{fig:cache_trans}).
%Because of the non-inclusive LLC, the lifetime of a cache line is so different from previous CPUs with inclusive LLC.
When a core accesses a fresh memory address (the core is called \textit{requesting core}), both L1 and L2 cache would miss. The memory sub-system of the requesting core computes a LLC slice ID from the memory address, to find the CHA of the core containing the LLC slice (called \textit{target core}). Then a read transaction is sent to the CHA of the target core (Step \textcircled{1}), which replies to the requesting core and updates the \textit{directory} accordingly.
Since the line is not cached in the LLC nor another core, CHA will ask IMC to fetch the line from memory (Step \textcircled{2}), and IMC will send the line directly to the requesting core (Step \textcircled{3}). The fresh cache line will be inserted into the L2 of the requesting core, but LLC will be kept untouched, because it is non-inclusive. When the requesting core runs out of L2 cache, it will follow a pseudo-LRU policy~\cite{yan2019attack} to evict a line from L2 to LLC (or dropped depending on the core's policy) (Step \textcircled{4}). Such an evicted line will be accessed from LLC the next time (Step \textcircled{5}). As the same cache line could persist in the cache of different cores, to avoid accessing the stale data, Intel employs MESIF (Modified, Exclusive, Shared, Invalid, Forward) protocol~\cite{mesh_mesif} for \textit{cache coherence}, which happens at Step \textcircled{2}.
%\zl{need to check with you again.}
To be noticed is that all the five steps leverage mesh to deliver transaction messages as well as the cache lines. This is critical to our attack as it exploits the statistics of the transactions to infer the victim's secret. In Section~\ref{subsec:path}, we further classify the traffic on the mesh.

\section{Fast Modular Exponentiation Algorithm}
\label{appendix:rsa}

The pseudo-code of RSA's fast modular exponentiation algorithm is shown in Algorithm~\ref{alg:rsa}.

\label{sec:mod_exp}
\begin{algorithm}[h]
\SetAlgoLined
\SetKwInput{KwInput}{Input}
\SetKwInput{KwOutput}{Output}
\KwInput{$m, d, n$}
\KwOutput{$res: m^d \; mod \; n$}

 $res = 1$\;
 \While{$d > 0$}{
  \If{$d \; mod \; 2 \; != 0 $}{
    $res = (res * m) \; mod \; n$
   }
   $d = d >> 1$\;
   $m = m^2 \; mod \; n$
 }
 \caption{Fast modular exponentiation algorithm
 %\zl{missing return statement [RA];In this style, return is placed at output.}
 }\label{alg:rsa}
\end{algorithm}

\section{\texttt{check\_conflict} and \texttt{find\_ev}}
\label{appendix:routines}

According to \cite{liu2015last, yan2019attack}, The \texttt{check\_conflict} function checks if an address $x$ conflicts with a set of addresses $U$, by checking if $x$ is evicted when traversing $x$ followed by $U$. If $x$ is evicted by $U$, it indicates $U$ conflicts with $x$, otherwise, it does not. \cite{yan2019attack} adapted this function to CPUs with non-inclusive LLC, by pushing all lines in a L2 set to LLC before measuring accessing time of $x$, which reduces false positives and negatives.

According to~\cite{yan2019attack}, the \texttt{find\_ev} function tries to find a minimal $EV$ within a given set of addresses $CS$. It starts by randomly picking out an address $x$ from $CS$ and assigning the rest addresses in $CS'$. It then repeatedly deletes addresses from $CS'$ except those addresses making $CS'$ no longer conflict with $x$. Those addresses should be in the $EV$. $EV$ could be further extended by picking out those addresses in $CS$ but conflict with $EV$.

\section{Layout of Xeon Scalable 8175}
\label{appendix:xs8175}

Table~\ref{tab:possible_pos} and~\ref{tab:cha_cpu_pos} show the reverse-engineered layout of Xeon Scalable 8175.

\begin{table}[h]
  \centering
  \begin{tabular}{|c|c|c|c|c|c|}
     \hline
     UPI  & PCIE & PCIE  & RLINK & UPI2 & PCIE \\ \hline
     0    & \cellcolor[gray]{0.8}4 & 9  & 14 & 19 & \cellcolor[gray]{0.8}24 \\ \hline
     IMC0 & 5 & 10 & 15 & 20 & IMC1 \\ \hline
     \cellcolor[gray]{0.8} 1    & 6 & 11 & 16 & 21 & 25 \\ \hline
     2    & 7 & 12 & 17 & 22 & 26 \\ \hline
     3    & 8 & 13 & 18 & 23 & \cellcolor[gray]{0.8} 27 \\ \hline
   \end{tabular}
  \caption{The disabled tiles of Xeon Scalable 8175 CPU. Gray cell indicates the tile is disabled, including its core, CHA, SF and LLC. The number in each cell is the ID of tile. The tiles without numbers like ``UPI'' do not have cores.}\label{tab:possible_pos}
\end{table}

\begin{table}[h]
  \centering
  \begin{tabular}{|c|c|c|c|c|c|}
     \hline
     UPI  & PCIE & PCIE  & RLINK & UPI2 & PCIE \\ \hline
     0, 0    & \cellcolor[gray]{0.8} & 7, 19  & 12, 3 & 17, 16 & \cellcolor[gray]{0.8} \\ \hline
     IMC0 & 3, 18 & 8, 2 & 13, 15 & 18, 10 & IMC1 \\ \hline
     \cellcolor[gray]{0.8}    & 4, 1 & 9, 14 & 14, 9 & 19, 22 & 22, 11 \\ \hline
     1, 12    & 5, 13 & 10, 8 & 15, 21 & 20, 5 & 23, 23 \\ \hline
     2, 6    & 6, 7 & 11, 20 & 16, 4 & 21, 17 & \cellcolor[gray]{0.8}  \\ \hline
   \end{tabular}
  \caption{Layout of Xeon Scalable 8175 CPU. Gray cell indicates the tile is disabled. The two numbers in each cell indicates the ID of CHA and core respectively. }\label{tab:cha_cpu_pos}
\end{table}

\section{Psuedo-code of covert channel}
\label{appendix:covert}

Algorithm~\ref{alg:covert} shows the pseudo-code of the covert-channel sender.

\begin{algorithm}[h]
\SetAlgoLined
\SetKwInput{KwInput}{Input}
\SetKwInput{KwOutput}{Output}
\KwInput{$T, bits, EV$}

\For{bit in bits}{
    ts = time()\;
    \eIf{bit == 1}{
        \While{time() $\leq$ ts + $\frac{T}{2}$}{
        access(EV)\;}
    }{
        \While{time() $\leq$ ts + $\frac{T}{2}$}{NOP\;}
    }
}

  \caption{The sender program used to test channel capacity}\label{alg:covert}
\end{algorithm}

\section{Selection of the Key Path}
\label{appendix:access}

If the attacker program does not run in the virtualized environment, it can learn the core ID of the victim, can select the key path.
Assuming the attack is running on Linux, the attacker can query a file \texttt{/proc/PID/stat} to obtain the process ID of the victim's program, and uses \texttt{ps -o psr} with the process ID as input~\cite{linux_ps}, to learn the core ID. The file and command are open to non-root users.
After that, the attacker maps the ID to the inferred CPU layout, compute $S$ of the four directions for T1-T3, and sends the probe packets.
Here we give an example about how to choose an attack path. Assume the victim is running at (5, 13) of Table~\ref{tab:cha_cpu_pos}, the attacker can bind a thread to core 8 (10, 8) and then probe CHA 23 (23, 23). In this case, $\frac{18}{23}$ of the mesh traffic sent to the victim core is supposed to contend with the attacker's traffic (the east and west traffic share bandwidth).
To increase the probabilities of contention, the attacker could occupy multiple cores and probe multiple CHAs. For example, she can also occupy core 21 (15, 21) and access CHA 22 (22, 11) at the same time. Yet, our evaluations suggests one path is enough.

\section{Analyzing the sequence of Sliding Window RSA}
\label{appendix:sw_seq}

\begin{figure}
    \centering
    \includegraphics[width = 0.45\textwidth]{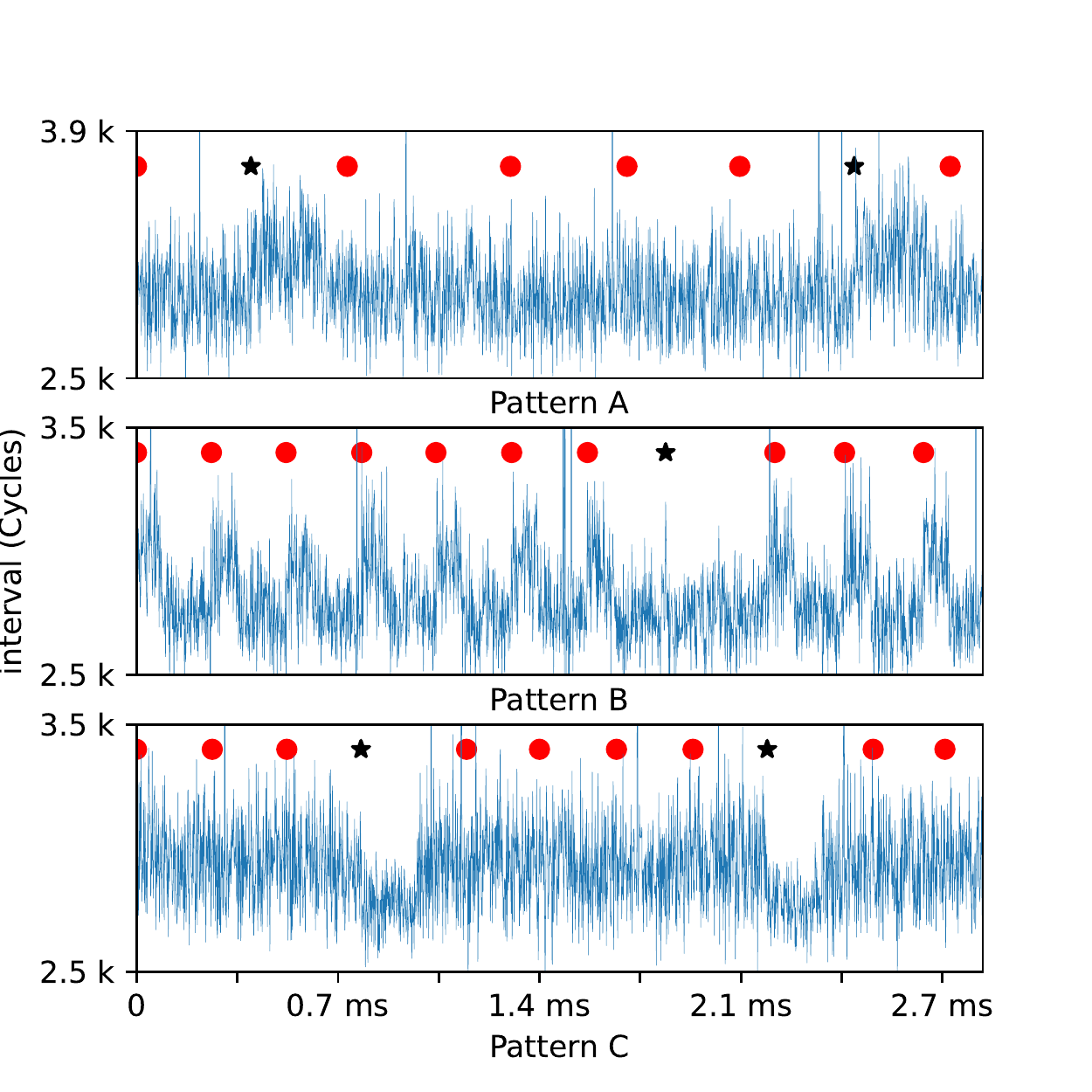}
    \caption{Interval Sequence collected for sliding window RSA. Black stars represent ground truth positions of mul, and red dots represent the positions of sqr. }
    \label{fig:sliding_wave}
\end{figure}

Figure~\ref{fig:sliding_wave} shows three different patterns. For Pattern A, each \texttt{mul} incurs an obvious rise, while for pattern C, each \texttt{mul} incurs a discernable pit. For pattern B, each \texttt{sqr} incurs a rise. Take pattern C as an example, the attacker needs only firstly deduce the positions of \texttt{mul}s, then can she know how many \texttt{sqr}s are there between consecutive \texttt{mul}s, according to the interval between consecutive \texttt{mul}s.

The method of processing the patterns is similar to that in Section~\ref{subsec:toy}, \ie, clipping first and then smoothing. But each pattern requires customized parameters. We adjusted three different parameters to handle them respectively. We firstly check if the pattern is obvious enough and then identify which category an observation falls into. Then we use sequence recovery with corresponding parameters accordingly.

\end{document}